\documentclass[hyper,11pt,letterpaper]{JHEP3}

\usepackage{graphicx, amsmath,amssymb,calrsfs,mathrsfs} 

\newtheorem{example}{Example}

\newcommand{\mc}[1]{\mathscr{#1}}
\newcommand{\rv}[1]{\boldsymbol{\mathsf{#1}}}
\newcommand{\ra}{\rightarrow}

\newcommand{\vev}[1]{\left\langle #1\right\rangle}

\title{Universal Nonlinear Filtering Using Feynman Path Integrals I: The Continuous-Discrete Model with Additive Noise}

\author{Bhashyam Balaji\\
Radar Systems Section,\\
Defence Research and Development Canada, Ottawa,\\
3701 Carling Avenue, \\
Ottawa ON K1A 0Z4 Canada\\
Email: Bhashyam.Balaji@drdc-rddc.gc.ca\\
}

\abstract{
The continuous-discrete filtering problem requires the solution of a partial differential equation known as the Fokker-Planck-Kolmogorov forward equation (FPKfe). In this paper, the path integral formula for the fundamental solution of the FPKfe is derived and verified for the general additive noise case (i.e., explicitly time-dependent state model and with state-independent rectangular diffusion vielbein). The solution is universal in the sense that the initial distribution may be arbitrary. The practical utility is demonstrated via some examples.
}


\begin{document}

\section{Introduction}
In a wide variety of applications, the evolution of a state, or a signal of interest, is described by a stochastic dynamical model. That is, the state of the system is described by a noisy version of a deterministic dynamical system termed the state model. If the state process is continuous (discrete), then the dynamics is governed by a system of first-order differential equations (difference equations) in the state variable ($\rv{x}(t)$) with an additional contribution due to random noise ($\rv{\nu}(t)$).  The noise in the state model is referred to as the signal noise. If the noise is Gaussian (or more generally,  multiplicative Gaussian) the state process is a Markov process.

However, in many applications the signal model cannot  be directly observed. Instead, what is measured is a nonlinearly related stochastic process ($\rv{y}(t)$) called the measurement process. The measurement process can often by modelled as yet another stochastic dynamical system called the measurement model. That is, the observations, or measurements, are drawn from a different system of noisy first order differential (difference) equations. The noise in the measurement dynamical system is referred to as measurement noise.  The nonlinear filtering problem is to estimate the state of a stochastic dynamical system, given the observations of a related stochastic measurement process. For an excellent discussion of the subject, see Jazwinski\cite{A.H.Jazwinski1970}. 

The conditional probability density function of the state parameters, given the observations,  is the complete solution of the filtering problem. It contains all the probabilistic information about the state process that is in the measurements and the initial condition. This is the Bayesian approach; i.e., the  \textit{a priori} initial data  about the signal process contained in the initial probability distribution of the state($u(t_0,x)$) is incorporated in the solution. Given the conditional probability density, optimality may be defined under various criteria. Usually, the conditional mean, which is the least mean-squares estimate, is studied due to its richness in results and mathematical elegance. The solution of the optimal nonlinear filtering problem is termed universal, if the initial distribution can be arbitrary. 

The signal and measurement processes may be discrete-time or continuous-time stochastic processes and there is  a different type of filtering problem for each possible combination of continuous (discrete) signal and measurement processes. In this paper, the measurement process is described by a discrete-time stochastic process and the underlying signal process is a continuous-time stochastic process. 

The linear filtering problem, i.e., where the state and measurement processes are linear, was investigated by Kalman and Bucy\cite{R.E.Kalman1960}, \cite{R.E.KalmanR.S.Bucy1961}. For a modern review of the subject, see \cite{T.KailathandA.H.SayedandB.Hassibi2000}. The Kalman filter has been successfully applied to a large number of problems.

In spite of its enormous success, the Kalman filter suffers from some major limitations. From a Bayesian perspective, the simplicity of the Kalman filter for the linear state model is because it is based on the assumption that the initial distribution is Gaussian. Then, the probability distribution function is completely characterized by the  mean and covariance matrices\footnote{Such a simple characterization of the probability distribution and solution is possible only for a limited class of filtering problems called  finite dimensional filters.}. The Kalman filter is still optimal if the initial distribution is not Gaussian only under certain criteria, such as minimum variance, but not under an arbitrary loss function. In other words, the Kalman filter is not a universal optimal filter even when the filtering problem is linear. In the general nonlinear case, the filter state is infinite dimensional, namely, the whole conditional probability distribution function, and the Kalman filter cannot be optimal and will fail, although it may still work quite well if the nonlinearity is mild enough. 

It is well known that the evolution of the probability distribution of the state variable is governed by the Fokker-Planck-Kolmogorov equation (FPKfe). The solution of the continuous-discrete filtering problem also requires the solution of a FPKfe, since the probability density between measurements evolves according to the FPKfe (see, for instance, \cite{A.H.Jazwinski1970}). The FPKfe is a linear, parabolic, partial differential equation(PDE)\footnote{Sometimes in the literature the term `nonlinear FPKfe' is used to denote the FPKfe corresponding to a state process with a nonlinear drift. However, this FPKfe is still a linear, parabolic PDE.}. The literature on methods to solve the FPKfe type of equations is vast. In a broad sense, there are three main types of methods: finite difference methods \cite{Thom'ee90,Marchuk90}, spectral methods \cite{C.CanutoandM.Y.HussainiandA.QuarteroniandT.A.ZangJr.2006} and finite element methods\cite{V.Thom'ee1997}.  

However, there are several difficulties in solving partial differential equations. A na\"ive discretization may not be convergent; i.e., the approximation error may not vanish as the grid size is reduced. Alternatively, when the discretization spacing is decreased, it may tend to a different equation; i.e., be inconsistent. Furthermore, it may be that the numerical method is unstable. Finally, since the solution of a FPKfe is a probability density, it must be positive.  This may not be guaranteed by the discretization of the FPKfe.  

A different approach was taken by Shing-Tung Yau and Stephen Yau in \cite{S.-T.YauS.S.-TYau1996}. A more general equation than the FPKfe, one which arises in the continuous-continuous nonlinear filtering problem with additive noise (and with no explicit time dependence), was studied in \cite{S.-T.YauS.S.-TYau1996}. It was shown that the formal solution may be written as an ordinary, but somewhat complicated, multi-dimensional integral, where the integrand is an infinite series. In addition, they also presented an estimate of the time needed for the solution to converge to the true solution.  Since this FPKfe is independent of the measurements, it can be solved off-line. 

In this paper, it is shown that the fundamental solution of the FPKfe, and hence the continuous-discrete filtering problem, can be solved in terms of a Feynman path integral  for the general additive noise state model. Specifically, the transition probability probability density is expressed as a Feynman path integral. Thus, the solution of the FPKfe is reduced to the computation of a path integral. Although the path integral is the limit of a large-dimensional integral, it shall be seen that the path integrand is considerably simpler than the integrand in \cite{S.-T.YauS.S.-TYau1996} (e.g., no convolution or infinite series).

Path integrals first arose in the study of quantum physics. Inspired by an observation/remark of Dirac\cite{P.A.M.Dirac1933, P.A.M.Dirac1982}, Feynman discovered the path integral representation of quantum physics \cite{R.P.FeynmanandA.R.Hibbs1965,Feynman1948}. Until the mid 1960s, path integrals were thought to be merely a curiosity, and not a serious alternative to the operator methods in quantum physics. The importance of path integrals grew when Fadeev and Popov used the path integrals to derive Feynman rules for the non-Abelian gauge theories. Its status was further enhanced when 't Hooft and Veltman used path integral methods to prove that the Yang-Mills theories were renormalizable. Since then, the path integral formulation of quantum field theory has led to extraordinary insights into quantum field theory, such as renormalization and renormalization group, anomalies, skyrmions, monopoles, instantons, and supersymmetry (see, for instance, the freely available text \cite{siegel-1999}). Much of the advance in the past 40 years in particle theory would not have been possible without the path integral. The path integral has also led to numerous insights into mathematics, such as the quantum field theory representation of the knots invariants, instantons and monopoles in supersymmetric gauge theories and the Donaldson-Seiberg-Witten invariants of four-manifolds, and more recently, S-duality in supersymmetric gauge theories and the Laglands program in mathematics. For a rigorous account of path integral methods, see \cite{JamesGlimmandArthurJaffe1987}.

The path integral representation is especially interesting from a computational point of view. A classic example is lattice quantum chromodynamics (QCD) which is the lattice formulation of the quantum field theory of strong interactions. Observables in quantum field theory require path integration over quantum fields whose dynamics is described by highly nonlinear infinite-dimensional systems. The path integral method has been used to perform the nonperturbative computation of quantities in quantum chromodynamics (see, for instance, \cite{H.J.Rothe2005}). The comparison to experimental results, even at coarse resolutions, with lattice computations is excellent. 

The following point needs to be emphasized to readers familiar with the discussion of standard filtering theory. In filtering theory literature, it is the Feynman-K\v ac formalism that is used. The Feynman-K\v ac formulation is rigorous and has led to several rigorous results in filtering theory. However, in spite of considerable effort it has not been proven to be useful in the development of practical algorithms. It also obscures the physics of the problem. In contrast, it is shown that the Feynman path integral leads to formulas that are eminently suitable for numerical implementation. It also provides a simple and clear physical picture. Finally, the theoretical insights provided by the Feynman path integral are highly valuable, as evidenced by numerous examples mentioned above. 

The path integral solution of the FPKfe has been studied in various branches of theoretical physics for different purposes. This is because the FPKfe (and the Langevin equation) is a fundamental equation of statistical physics. A textbook discussion of the FPKfe is presented in  \cite{H.Risken1999}, and \cite{JeanZinn-Justin2002}.   The results derived in this paper generalize some of the results in \cite{JeanZinn-Justin2002}. Specifically, the path integral formula for the fundamental solution is derived and independently verified for the case when the drift in the state model is explicitly time-dependent, the diffusion vielbein is an explicitly time dependent rectangular matrix, and when the noise process is colored with a time dependent covariance matrix. In addition, the discussion here improves upon the discussion in \cite{JeanZinn-Justin2002} by clarifying important and subtle aspects of the derivation of the path integral formulas and by providing more details so that this powerful technique is accessible to a wider audience.  

The outline of this paper is as follows. In Section \ref{sec:revcdfilt}, the Langevin equation, the corresponding Fokker-Planck-Kolmogorov forward equation, and the role the FPKfe plays in solving the continuous-discrete filtering problem, are reviewed. In Section \ref{sec:Prelims},  the meaning of the Gaussian path integral measure is clarified and some important properties of the Langevin equation are reviewed. In Section \ref{sec:pathinteg1}, two derivations are presented of the path integral formula for the fundamental solution for the simpler case; i.e., where the drift and diffusion vielbein are not explicitly time dependent and the noise is white and additive. This is likely to be the case of most interest in practise. In Section \ref{sec:pathinteg2}, this is generalized to the case when the drift and diffusion vielbein are explicitly time dependent and the additive noise is colored. In Section \ref{sec:Verification}, it is independently verified that the derived path integral formula does indeed satisfy the FPKfe. In the following section, some numerical examples are presented. This is followed by a discussion of the physical meaning of the path integral formula and a comparison to the solution with that obtained by S-T. Yau and S. S-T. Yau.

In the appendices, the tools used in the derivations are reviewed including Gaussian integration, the imposition of delta function constraints as an integral, and aspects of functional calculus. Although most of these topics are discussed in modern quantum field theory texts, it has been included in order to make the paper completely self-contained.

\section{Continuous-Discrete Filtering: A Review }\label{sec:revcdfilt}

\subsection{The Fokker-Planck-Kolmogorov forward Equation}

In continuous-discrete filtering theory, the state model is given by the stochastic differential equation  of the form
\begin{align}\label{eq:modelStrat000}
	d\rv{x}(t)=f(\rv{x}(t),t)dt+e(\rv{x}(t),t)d\rv{v}(t).
\end{align}
Here $\rv{x}(t)$ and $f(\rv{x}(t),t)$ are $n$-dimensional column vectors, $e(\rv{x}(t),t)$ is an $n\times p$ matrix and $\rv{v}(t)$ is a $p-$dimensional Wiener process column vector. In physics literature, Equation \ref{eq:modelStrat000} is written as follows:
\begin{align}
\label{eq:statemodelStratanovich00}
\frac{d\rv{x}(t)}{dt}=f(\rv{x}(t),t)+e(\rv{x}(t),t)\rv{\nu}(t).
\end{align}
The stochastic process $\rv{\nu}(t)$ is assumed to be Gaussian with zero mean and ``$\delta-$correlated'', i.e.,
\begin{align}
\label{eq:GaussMeanCov}
	\vev{\rv{\nu}(t)}&=0,\\
	\vev{\rv{\nu}(t)\rv{\nu}^T(t')}&=Q(t)\delta(t-t'),  \nonumber
\end{align}
where $Q(t)$ is a $p\times p$ covariance matrix. The quantity $f$ is referred to as the (signal model) drift and $e$ as the diffusion vielbein, and the quantity $eQe^T$ is referred to as the diffusion matrix. The stochastic processes are represented in bold and one of its samples by its corresponding plain symbol. In the interest of clarity, the tensor indices are suppressed in this subsection. The expectation with respect to Gaussian noise is denoted by angular brackets. Equation \ref{eq:statemodelStratanovich00} is also referred to as the Langevin equation with multiplicative noise and $\delta-$correlated Langevin force. When $e=e(t)$, i.e., the diffusion vielbein is independent of the state variable,  the noise in the Langevin equation is termed additive noise. 

Consider an ensemble of dynamical systems with state variables evolving according to the Langevin equation. Due to random noise, each system leads to a different vector $x(t)$ that depends on time. Although only one realization of the stochastic process is observed, it is meaningful to consider an ensemble average. For fixed times $t=t_i,i=1,2,\dots,r$, the probability density of finding the random vector $\rv{x}(t)$ in the ($n-$dimensional) interval $x_i\le \rv{x}(t_i)\le x_i+dx_i(1\le i\le r)$ is given by
\begin{align}
	W_r(t_r,x_r;\cdots;t_1,x_1)=\vev{\prod_{i=1}^r\delta^n(\rv{x}(t_i)-x_i)}, 
\end{align}
where $x_i$ is an $n-$dimensional column vector. The complete information on the random vector $\rv{x}(t)$ is contained in the infinite hierarchy of such probability densities. The quantity of interest here is the conditional probability density\footnote{Unless otherwise specified, all integration variables are from $-\infty$ to $\infty$.}
\begin{align}\label{eq:CondProbDens}
	P(t_r,x_r|t_{r-1},x_{r-1},\ldots;t_1,x_1)&=\vev{\delta^{n}(\rv{x}(t_r)-x(t_r))}|_{x(t_{r-1})=x_{r-1},\dots,x(t_1)=x_1},\quad x(t_r)\equiv x_r\\ \nonumber
	&=\frac{W_r(t_r,x_r;\ldots;t_1,x_1)}{\int W_r(t_r,x_r;\ldots;t_1,x_1)d^nx_r}.
\end{align}

The process described by the Langevin equation with $\delta-$correlated Langevin force is a Markov process; i.e., the conditional probability density depends only on the value at the immediate previous  time:
\begin{align}
	P(t_r,x_r|t_{r-1},x_{r-1};\ldots;t_1,x_1)=P(t_n,x_n|t_{n-1},x_{n-1}).
\end{align}
This implies that 
\begin{align}
	W_r(t_r,x_r;\ldots;t_1,x_1)=\left[\prod_{i=1}^{r-1}P(t_{i+1},x_{i+1}|t_i,x_i)\right]W_1(t_1,x_1).
\end{align}
Hence, the complete information for a Markov process is contained in the transition probability densities
\begin{align}\label{eq:MarkovProp01}
	P(t_2,x_2|t_1,x_1)=\frac{W_2(t_2,x_2;t_1,x_1)}{W_1(t_1,x_1)}. 
\end{align}

From the identity
\begin{align}\label{eq:ChapKol00}
	W_2(t_3, x_3;t_1,x_1)=\int W_3(t_3,x_3;t_2,x_2;t_1,x_1)\{d^nx_2\},
\end{align}
and using the Markov property Equation \ref{eq:ChapKol00} may be written as
\begin{align}
	P(t_3,x_3|t_1,x_1)W_1(t_1,x_1)=\int P(t_3,x_3|t_2,x_2)P(t_2,x_2|t_1,x_1)W_1(t_1,x_1)\{d^nx_2\}. 
\end{align}
Since $W_{1}(t_{1},x_{1})$ is arbitrary, the Chapman-Kolmogorov equation results:
\begin{align}
	P(t_3,x_3|t_1,x_1)=\int P(t_3,x_3|t_2,x_2)P(t_2,x_2|t_1,x_1)\{d^nx_2\}. 
\end{align}
This is also referred to as the Chapman-Kolmogorov semi-group property of the transition probability density. This property plays a fundamental role in arriving at the path integral formula. 

It can be shown that the state probability distribution function $p(t,x)=\int P(t,x|t'x')p(t',x')\{d^nx'\}$ satisfies the Fokker-Planck-Kolmogorov forward equation (FPKfe) (see for instance, \cite{H.Risken1999}): 
\begin{align}
	\frac{\partial p}{\partial t}(t,x)&=-\sum_{i=1}^n\frac{\partial }{\partial x_i}\left[f_i(x(t),t)p(t,x)\right]+\frac{1}{2}\sum_{i,k=1}^n\sum_{b=1}^p\frac{\partial }{\partial x_i}\left[ (e(x(t),t)Q(t))_{ib}\frac{\partial}{\partial x_k}(e^T_{bk}(x(t),t)p(t,x)) \right],\\ \nonumber
	&\equiv{\mc{L}}p(t,x).
\end{align}

The diffusion matrix is clearly a symmetric matrix and is also positive semi-definite: $D_{ij}a_ia_j\ge0$. Often, $D_{ij}$ is assumed to be positive-definite, in which case the inverse of the diffusion matrix exists. An important property of the FPKfe is that it is a continuity equation.

\subsection{Continuous-Discrete Filtering: Model and Solution}

In this paper, it is assumed that the dynamics is given by a continuous-time process and observations are samples of a discrete-time process. The continuous-time state model is described by Equation \ref{eq:modelStrat000}. 
The measurement model is described by the  following discrete-time stochastic process
\begin{align}
	\rv{y}(t_k)=h(\rv{x}(t_k),t_k)+\rv{w}(t_k),\quad k=1,2,\dots, \quad t_{k+1}>t_k\geq0,
\end{align}
where $y(t)\in\mathbb{R}^{m\times1}$, $h\in\mathbb{R}^{m\times1}$ and the noise process is described by  $\rv{w}(t)\sim N(0,R(t))$. As discussed below, the Gaussian assumption in the measurement model can be relaxed.

Let the initial distribution be $\sigma_0(x)$ and let the measurements be collected at time instants $t_1, t_2,\ldots, t_k,\ldots$. We use the notation $Y(\tau)=\left\{ y(t_l):t_0<t_l\le\tau \right\}$. Prior to incorporating the measurements, the state evolves according to the FPKfe; i.e., 
\begin{align}
	\frac{\partial p}{\partial t}(t,x|Y({t_0}))&={\mc{L}}(p(t,x|Y({t_0}))), \quad t_0<t\leq t_1,\\ \nonumber
	p(t_0,x|Y({t_0}))&=\sigma_0(x).
\end{align}
This is the `prediction' step.

From Bayes' rule (and using $p(t_i,x|Y(t_i))=p(t_i,x|y(t_i),Y(t_{i-1}))$ at observation $t_1$, the `corrected' conditional density at time $t_1$ is  
\begin{align}
	p(t_1,x|Y(t_1))=\frac{p(y(t_1)|x)p(t_1,x|Y(t_0))}{\int p(y(t_1)|\xi)p(t_1,\xi|Y(t_0))\{d^n\xi\}}.
\end{align}
This is then the initial condition of the FPKfe for the next prediction step which results in
\begin{align}
	\label{eq:CondDiscStep2}
	\frac{\partial p}{\partial t}(t,x|Y(t_1))&=\mc{L}(p(t,x|Y(t_1))),\quad t_1<t\le t_2,&\quad(\text{Prediction Step}),\\ \nonumber
	p(t_2,x|Y(t_2))&=\frac{p(y(t_2)|x)p(t_2,x|Y(t_1))}{\int p(y(t_2)|\xi)p(t_2,\xi|Y(t_1))\{d^n\xi\}},&\quad(\text{Correction Step}),
\end{align}
and so on. Thus, at observation at time $t_k$, the conditional density is given by 
\begin{align}
	p(t_k,x|Y(t_k))=\frac{p(y(t_k)|x)p(t_k,x|Y(t_{k-1}))}{\int p(y(t_k)|\xi)p(t_k,\xi|Y(t_{k-1}))\{d^n\xi\}},
\end{align}
where $p(y(t_k)|x)$ is given by 
\begin{align}\label{eq:MeasGaussCorr}
	p(y(t_k)|x)=\frac{1}{\left( (2\pi)^m\det R(t_k) \right)^{1/2}}\exp\left\{ -\frac{1}{2}(y(t_k)-h(x(t_k),t_k))^T(R(t_k))^{-1}(y(t_k)-h(x(t_k),t_k)) \right\},
\end{align}
and $p(t_k,x|Y(t_{k-1}))$ is given by the solution of the PDE
\begin{align}	
	\frac{\partial}{\partial t}p(t,x|Y(t_{k-1}))&={\mc{L}}(p(t,x|Y(t_{k-1}))),\quad t_{k-1}\le t<t_k, 
\end{align}
with initial condition $p(t_{k-1},x|Y({t_{k-1}}))$. Note that the measurement model noise need not be Gaussian; the Gaussian assumption merely specifies the form of $p(y(t_{k})|x)$ to be given by Equation \ref{eq:MeasGaussCorr}.

Thus, the non-trivial part of continuous-discrete filtering is the solution of a FPKfe. Note that the FPKfe is independent of measurements, and hence can be computed off-line. Furthermore, observe that the complete information is in the transition probability density which also satisfies the FPKfe except with a $\delta-$function initial condition (known as the fundamental solution, or kernel, of the FPKfe):
\begin{align}\label{eq:FPKfePDE}
	\frac{\partial}{\partial t}p(t,x|\tau,y)&={\mc{L}}(p(t,x|\tau,y)),\qquad t>\tau,\\ \nonumber
	p(t,x|t,y)&=\delta(x-y).
\end{align}
 This is because from the transition probability density the probability for an arbitrary initial condition can be computed as follows:
\begin{align}\label{eq:FPKfearbInitCond}
	p(t_n,x)=\int p(t_n,x|t_{n-1},x')p(t_{n-1},x')\{d^nx'\}.
\end{align}
Thus, it is sufficient to solve the FPKfe for $p(t,x|\tau,y)$ and use the above to compute $p(t_k,x|Y(t_{k-1}))$ using
\begin{align}
	p(t_k,x|Y(t_{k-1}))=\int p(t_k,x|t_{k-1},x')p(t_{k-1},x'|Y(t_{k-1}))\{d^nx'\}.
\end{align}

Alternatively, a convenient set of basis functions may be used. Then, the evolution of each of the basis functions under the  FPKfe follows from Equation \ref{eq:FPKfearbInitCond}. Since the basis functions  are independent of measurements, the computation may be performed off-line\cite{LiangYau2004}. Note that the solution is universal; ie., the initial distribution can be arbitrary. 

In conclusion, the solution of Equation \ref{eq:FPKfePDE} is equivalent to the solution of the universal optimal nonlinear filtering problem. A solution in terms of ordinary integrals was presented in \cite{S.-T.YauS.S.-TYau1996}.  Here, the solution in terms of Feynman path integrals is presented. 

\section{Some Preliminary  Remarks}\label{sec:Prelims}

The reader may refer to the appendices for the derivation of the results used here.

\subsection{Gaussian Path Integral Measure} 

The Gaussian noise process $\rv{\nu}(t)$, defined in Equation \ref{eq:GaussMeanCov}, may be represented by a path integral measure 
\begin{align}
\label{eq:GaussMeasure}
[d\rho(\nu(t))]=[\mc{D}\nu(t)]\exp\left[ -\frac{1}{2}\int dt\sum_{a,b=1}^p\nu_a(t)\left(Q^{-1}(t)\right)_{ab}\nu_b(t) \right], \quad \nu\in\mathbb{R}^{p\times1},
\end{align}
where $\nu(t)$ is a real vector for each $t$. The meaning of Equation \ref{eq:GaussMeasure} is explained next.

Recall  that the probability measure of a $n-$dimensional Gaussian vector $\rv{\nu}$ with zero mean and covariance matrix $Q$ is 
\begin{align}
	\left\{\frac{1}{\sqrt{\det Q}}\left( \frac{1}{2\pi} \right)^{p/2}\left\{d^n\nu\right\}\right\}\exp\left(-\frac{1}{2}\sum_{a,b=1}^p\nu_aQ_{ab}^{-1}\nu_b\right).
\end{align}
The first moment or mean is  
\begin{align}
\label{eq:Norm00}
\vev{\rv{\nu}}&=\int \left\{\frac{1}{\sqrt{\det Q}}\left(\frac{1}{2\pi} \right)^{p/2}\left\{d^n\nu\right\}\right\} \nu_c\exp\left(-\frac{1}{2}\sum_{a,b=1}^p\nu_aQ_{ab}^{-1}\nu_b\right),\\ \nonumber &=0,
\end{align}
by virtue of the odd symmetry of the integrand. The second moment is easily seen to be (e.g., from discussion in Appendix \ref{sec:GaussianIntegrals})
\begin{align}
	\vev{\rv{\nu}_a\rv{\nu}_b}&=\frac{\int \left\{d^n\nu\right\}\nu_c\nu_d\exp\left(-\frac{1}{2}\sum_{a,b=1}^p\nu_aQ_{ab}^{-1}\nu_b\right)}{\int \left\{d^n\nu\right\}\exp\left(-\frac{1}{2}\sum_{a,b=1}^p\nu_aQ_{ab}^{-1}\nu_b\right)},\\ \nonumber
	&=Q_{ab}.
\end{align}
It is straightforward to generalize the above for the case of a sequence of random vectors which are uncorrelated with each other at different times:
\begin{align} 
	\label{eq:Norm02}
	&\int \left\{ \prod_{k=1}^N\frac{1}{\sqrt{\det(Q(t_k))}}\left( \frac{1}{2\pi} \right)^{n/2} \left\{d^n\nu(t_k)\right\}\right\}\nu_c(t_k)\nu_d(t_l)\\ \nonumber
	&\quad\times\exp\left(-\frac{1}{2}\sum_{k=1}^N\sum_{a,b=1}^p\nu_a(t_k)Q_{ab}^{-1}(t_k)\nu_b(t_k)\right)=Q_{ab}(t_k) \delta_{kl}.
\end{align}
Note that the noise may be colored. It is also important to note that had uncorrelatedness at different times not been assumed, the exponent would not be a sum of quadratic terms local in time. 

For the continuum limit, the following expression is of interest (in the notation of Appendix \ref{sec:GaussianIntegrals}, $\alpha(t)=\frac{1}{2}, \beta=\Delta t$)
\begin{align} 
	\label{eq:Norm02b}
	&\int \left\{\prod_{k=1}^N \frac{1}{\sqrt{\det(Q(t_k))}}\left( \frac{1}{2\pi\Delta t} \right)^{n/2} \left\{d^nx(t_k)\right\}\right\}x_c(t_k)x_d(t_l)\\ \nonumber
	&\quad\times\exp\left(-\frac{\Delta t}{2}\sum_{k=1}^N\sum_{a,b=1}^px_a(Q^{-1}(t_k))_{ab}x_b\right)=\frac{1}{\Delta t} Q_{ab}(t_k)\delta_{kl}.
\end{align}
The measure in the continuum limit (i.e., $\Delta t\ra0$) is formally written as
\begin{align}
	[d\rho(\nu(t))]\equiv\left[ {\mc{D}}\nu(t) \right]\exp\left(-\frac{1}{2}\int \sum_{a,b=1}^p\nu_a(t)(Q^{-1})_{ab}\nu_b(t)dt\right),
\end{align}
where 
\begin{align}
	\label{eq:NorMeasure00}
	[\mc{D} \nu(t)]=\lim_{N\ra\infty}\prod_{k=1}^N\left\{ \left( \frac{1}{2\pi\Delta t}\right)^{n/2}\frac{1}{\sqrt{\det Q(t_k)}}\left\{d^n\nu(t_k)\right\} \right\}.
\end{align}
Hence, in the continuum limit, the mean is easily seen to be 0 and Equation \ref{eq:Norm02b} becomes
\begin{align}
	\vev{\rv{\nu}_a(t)\rv{\nu}_b(t')}=Q_{ab}(t)\delta(t-t').
\end{align}

These results may also be directly derived using the methods of functional calculus(see Appendix \ref{sec:FuncCalc}). In particular, it is straightforward to verify that the mean is zero due to odd symmetry (setting $\alpha=1/2$)
\begin{align}
	\vev{\rv{\nu}_c(t)}&=\int[d\rho(\nu)]\nu_c(t),\\ \nonumber
	&=\int[\mc{D}\nu(t)]\nu_c(t)\exp{\left(-\frac{1}{2}\int dt\sum_{a,b=1}^p\nu_a(t)(Q^{-1}(t))_{ab}\nu_b(t)\right)},\\ \nonumber
	&=0,
\end{align}
and the variance is $Q(t)$:
\begin{align}
	\vev{\rv{\nu}_c(t)\rv{\nu}_d(t')}&=\int[d\rho(\nu)]\nu_c(t)\nu_d(t'),\\ \nonumber
	&=\int[d\nu(t)]\nu_c(t)\nu_d(t')\exp{\left(-\frac{1}{2}\int dt\sum_{a,b=1}^p\nu_a(t)(Q^{-1}(t))_{ab}\nu_b(t)\right)},\\ \nonumber
	&=\frac{\delta}{\delta b_c(t)}\frac{\delta}{\delta b_d(t)}\exp\left(-\frac{1}{2}\int dt' \sum_{a,b=1}^pb_a(t')Q_{ab}(t')b_b(t')\right)\Big|_{b=0},\\ \nonumber
	&=Q_{cd}(t)\delta(t-t'),
\end{align}
and higher-order moments can easily be written down using Wick's theorem. The results are in accord with the expectation that the measure represents a Gaussian process with these first two moments.

\subsection{Basic Properties of the Langevin Equation}

We conclude this section with some fundamental properties of the Langevin Equation.
\begin{itemize}
	\item Time Translation invariance: When the drift does not depend on time explicitly, i.e., $f(\rv{x}(t))$, then the process is time-translation invariant; i.e., does not depend on the origin of time. Alternatively, the time translation operator $\frac{\partial}{\partial t}$ commutes with the FPK operator $\mc{L}$:
		\begin{align}
			\left[ \frac{\partial}{\partial t},\mc{L} \right]\Phi=\frac{\partial}{\partial t}\mc{L}\Phi-\mc{L}\frac{\partial}{\partial t}\Phi=0,
		\end{align}
		where $\Phi$ is an arbitrary differentiable function.
	\item Locality: Langevin equation is a differential equation specifying the time evolution as a system of first order stochastic differential equations with all terms evaluated at the same time. In other words, the Langevin equation is local in time.
	\item Noise uncorrelated at different times: This is an important assumption which leads to the $\delta-$correlated in time property of the noise process. 
\end{itemize}
From the locality of the Langevin equation and the uncorrelatedness of noise at different times, it immediately follows that $P$ satisfies the Chapman-Kolmogorov semi-group property (Equation \ref{eq:ChapKol00}), an important property of Markov processes: 
\begin{align}
	P(t_3,x_3|t_1,x_1)&=\int P(t_3,x_3|t_2,x_2)P(t_2,x_2|t_1,x_1)\{d^nx_2\}.
\end{align}
This implies that the distribution $P$ is completely determined from its knowledge for small time intervals. It is important to note that if the Langevin equation was not local and/or the noise was not uncorrelated at different times, the Chapman-Kolmogorov semi-group property would not be valid\footnote{ However, if noise is correlated at different times, some non-Markovian processes (for instance, exponentially correlated) may be reduced to Markovian processes. }.

\section{Path Integral Formula I: Implicit Time Dependence}\label{sec:pathinteg1}

In this section, the noise is assumed to be additive, white, and Gaussian-distributed. The state model is given by
\begin{align}
\label{eq:LangAdd01}
d\rv{x}(t)=f(\rv{x}(t))dt+d\rv{v}(t), \quad \rv{x}(t),f(\rv{x}(t)),\rv{v}(t)\in\mathbb{R}^n, \quad Q_{ij}=\hbar_{\nu}\delta_{ij}.
\end{align}
Observe that the drift is not explicitly time-dependent, and the diffusion vielbein is proportional to the identity matrix with proportionality constant $\hbar_{\nu}$. In this instance, the It\^o and Stratanovich forms are the same. A sample of this stochastic process is the differential equation 
\begin{align}
	\dot{x}_i(t)=f_i(x(t))+\nu_i(t),\quad i=1,2,\dots,n,
\end{align}
where $\nu$ is the sample of a Gaussian noise process with $\delta-$correlated covariance matrix $\hbar_{\nu}I$.  
\subsection{Formal Derivation Using Functional Calculus}\label{sec:FormalPlain}

The transition probability density, $\vev{\delta\left[ \rv{x}(t)-x \right]}_{\rv{\nu}}|_{x(t_0)=x_0}$, is now derived using functional methods for state model given by the Langevin equation given by  Equation \ref{eq:LangAdd01}. From the path integral representation of the Gaussian noise in Equation \ref{eq:GaussMeasure}, it follows that
\begin{align}
	\vev{\delta^n\left[ \rv{x}(t)-x \right]}_{\rv{\nu}}|_{x(t_0)=x_0}&=\int\left[ d\rho(\nu) \right]\delta^n\left[ x(t)-x\right]|_{x(t_0)=x_0},\quad\dot{x}(t)=f(x(t))+\nu(t).
\end{align} 
The condition that $x(t)$ in the integrand satisfies the Langevin equation needs to be imposed. This can be done using the identity Equation \ref{eq:UnityFunctional} as follows:
\begin{align}
	P(t,x|t_0,x_0)&=\int\left[ \mc{D}\nu(t) \right]\exp\left( -\frac{1}{2\hbar_{\nu}}\sum_{i=1}^n\int_{t_0}^t \nu_i^2(t)dt \right)\\ \nonumber
	&\qquad\times\int\left[ \mc{D}\dot{x}(t') \right]J\delta^n(\dot{x}(t)-f(x(t))-\nu(t)),\\ \nonumber
	&\qquad\times\delta^n[x(t)-x]|_{x(t_0)=x_0},
\end{align}
where 
\begin{align}
	J=\prod_{i=1}^n\det\left( \frac{\delta}{\delta x_j(t')}[\dot{x}_i(t)-f_i(x(t))-\nu_i(t)] \right).
\end{align}
Interchanging the order of integration and then performing the trivial integration over $\nu(t)$ leads to 
\begin{align}
	P(t,x|t_0,x_0)=\int\left[ \mc{D}\dot{x}(t) \right]\exp\left( -\frac{1}{2\hbar_{\nu}}\sum_{i=1}^n\int_{t_0}^t\left(\dot{x}_i(t)-f_i(x(t)  \right)^2 dt\right)J\delta^n[x(t)-x]|_{x(t_0)=x_0}.
\end{align}
It remains to compute the Jacobian and convert the  path integral to one over $x(t)$. The Jacobian $J$ follows from noting that
\begin{align}
\label{eq:Jacobian00}
	\frac{\delta}{\delta x_j(t')}\left[ \dot{x}_i(t)-f_i(x(t))-\nu_i(t) \right]&=\left[ \delta_{ij}\frac{d}{dt}- \frac{\partial f_i}{\partial x_j}(x(t'))\right]\delta(t-t'),\\ \nonumber
	&=\frac{d}{dt}\left[ \delta_{ij}\delta(t-t')-\theta(t-t')\frac{\partial f_i}{\partial x_j}(x(t')) \right],  \nonumber
\end{align}
because
\begin{align}
	\frac{d}{dt}\theta(t-t')=\delta(t-t'),\quad\text{ and }\quad \frac{\delta x_i(t)}{\delta x_j(t')}=\delta_{ij}\delta(t-t').
\end{align}
The Jacobian of the transformation is then given by 
\begin{align}
	\label{eq:Jacobian01}
	J&\equiv\det\left( \frac{d}{dt}\left[ \delta_{ij}\delta(t-t')-\theta(t-t')\frac{\partial f_i}{\partial x_j}(x(t')) \right] \right),\\ \nonumber
	&=\det\left( \frac{d}{dt} \right)\det\left( \delta_{ij}\delta(t-t')-\theta(t-t')\frac{\partial f_i}{\partial x_j}(x(t')) \right),\\ \nonumber
	&=\mc{N}\det\left( \delta_{ij}\delta(t-t')-\theta(t-t')\frac{\partial f_i}{\partial x_j}(x(t') )\right),  
\end{align}
where $\mc{N}$ is an irrelevant constant. From the identity $\det A=\exp\left( \text{tr}\ln A \right)$, and using the identity in Equation \ref{eq:DetIdentity}, it follows that  
\begin{align}
	\label{eq:Jacobian02}
	\text{ln}J&=\ln\det\left[ \delta_{ij}\delta(t-t')-\theta(t-t')\frac{\partial f_i}{\partial x_j}(x(t') \right],\\ \nonumber
	&=-\theta(0)\sum_{i=1}^n\int_{t_0}^t dt\frac{\partial f_i}{\partial x_i}(x(t)),\\ \nonumber
	&=-\frac{1}{2}\sum_{i=1}^n\int_{t_0}^t\frac{\partial f_i}{\partial x_i}(x(t))dt.
\end{align}
Since the symmetric or Feynman convention is used here\footnote{The Stratanovich convention symmetrizes the argument of the diffusion vielbein, while the Feynman convention symmetrizes the argument of the drift.}, $\theta(0)=\frac{1}{2}$. It also ensures that the operations of averaging and differentiation commute (see, for instance, \cite{JeanZinn-Justin2002}). Also, 
\begin{align}\label{eq:Measxdot}
	\int\left[ \mc{D}\dot{x}(t) \right]\ra\int [\mc{D}x(t)]\det\left( \frac{\delta\dot{x}(t)}{\delta x(t)} \right)=\det\left(\frac{d}{dt}\delta(t-t')\right)\int\left[ \mc{D}x(t) \right]\ra\int[\mc{D}x(t)],
\end{align}
where the irrelevant constant $\det(d/dt)\delta(t-t')$ has been absorbed into the measure.

Combining these results leads to the following path integral formula:
\begin{align}\label{eq:PIformula}
	P(t,x|t_0,x_0)=\int_{x(t_0)=x_0}^{x(t)=x}[\mc{D}x(t)]\exp\left( -\frac{1}{2\hbar_{\nu}}\sum_{i=1}^n\int_{t_{0}}^t dt\left[\left( \dot{x}_i(t)-f_i(x(t)) \right)^2 +\hbar_{\nu}\frac{\partial f_i}{\partial x_i}(x(t))\right]\right).
\end{align}

\subsection{Derivation using the Chapman-Kolmogorov Semi-group property}

Consider the time evolution of the state in an infinitesimal time interval. Integrating the Langevin equation (Equation \ref{eq:LangAdd01}) in an infinitesimal time interval $(t,t+\epsilon)$ yields
\begin{align}\label{eq:DiscLangSimple}
	\rv{x}_i(t+\epsilon)=\rv{x}_i(t)+\epsilon f_i(\rv{x}(t))+\int_t^{t+\epsilon}\rv{\nu}_i(\tau)d\tau+O(\epsilon^{3/2}).
\end{align}
To order $\epsilon$, the continuum Langevin equation is equivalent to the discretized Langevin equation:
\begin{align}
	\rv{x}_i(t+\epsilon)=\rv{x}_i(t)+\epsilon f_i(\rv{x}(t))+\sqrt{\epsilon}\bar{\rv{\nu}}_i(t),
\end{align}
where
\begin{align}
	\sqrt{\epsilon}\bar{\rv{\nu}}_i(t)\equiv\int_t^{t+\epsilon}\rv{\nu}_i(\tau)d\tau.
\end{align}
From the properties of $\rv{\nu}_i(t)$, it follows that
\begin{align}
	\vev{\sqrt{\epsilon}\bar{\rv{\nu}_i}(t)}
	&=\int_t^{t+\epsilon}\vev{\rv{\nu}_i(\tau)}d\tau,\\ \nonumber
	&=0,
\end{align}
and
 \begin{align}
	 \epsilon\vev{\bar{\rv{\nu}}_i(t)\bar{\rv{\nu}}_j(t')}&=\vev{\int_t^{t+\epsilon}\int_{t'}^{t'+\epsilon}\rv{\nu}_i(\tau)\rv{\nu}_j(\tau')d\tau d\tau'},\\ \nonumber
	 &=\int_t^{t+\epsilon}\int_{t'}^{t'+\epsilon}\vev{\rv{\nu}_i(\tau)\rv{\nu}_j(\tau')}d\tau d\tau',\\ \nonumber
	 &=\int_t^{t+\epsilon}\int_{t'}^{t'+\epsilon}\hbar_{\nu}\delta_{ij}\delta(\tau-\tau')d\tau d\tau',\\ \nonumber
	&=
	\begin{cases}
		0&t\neq t',\\
		\epsilon \hbar_{\nu}\delta_{ij}&t=t'
	\end{cases}.
\end{align}
Therefore,
\begin{align} \label{eq:DiscNoise00} 
	\vev{\bar{\rv{\nu}}_i(t)}&=0,\\ \nonumber
	\vev{\bar{\rv{\nu}}_i(t)\bar{\rv{\nu}}_j(t')}&=\hbar_{\nu}\delta_{ij}\delta_{tt'},
\end{align}
which implies that $\bar{\rv{\nu}}(t)$ is also  a (discrete-time) Gaussian process.

Consider the quantity $P(t+\epsilon,x|t,x')$. The Fourier transform $\tilde{P}$ of $P$ w.r.t. $x$ is 
\begin{align}
	\tilde{P}(t+\epsilon,p|t,x')&=\int \left\{d^nx\right\}\exp\left({-i\sum_{i=1}^np_i x_i}\right)P(t+\epsilon,x;t,x'),\\ \nonumber
	&=\int \left\{ d^nx \right\}\exp\left(-i\sum_{i=1}^np_ix_i\right)\vev{\delta^n(\rv{x}(t+\epsilon)-x)}|_{\rv{x}(t)=x}.
\end{align}
The order of integration over $x$ and averaging are interchanged to yield
\begin{align}
\tilde{P}(t+\epsilon,p|t,x')&=\vev{\exp\left(-i\sum_{i=1}^np_i\rv{x}_i(t+\epsilon)\right)},\\ \nonumber
	&=\exp\left[ -i\sum_{i=1}^np_i(x_i'+\epsilon f_i(x')) \right]\vev{\exp\left[ -i\sqrt{\epsilon}\sum_{i=1}^np_i\bar{\rv{\nu}}_i(t)\right]}.
\end{align}
Now, 
\begin{align} 
	\vev{\exp\left(-i\sqrt{\epsilon}\sum_{i=1}^np_i\bar{\rv{\nu}}_i(t)\right)}&= \int\frac{1}{\sqrt{(2\pi\hbar_{\nu})^n}}\left\{ d^n\bar{\nu}(t) \right\}\exp\left(-i\sqrt{\epsilon}\sum_{i=1}^np_i\bar{\nu}_i(t)\right) \\ \nonumber
	&\qquad\times\exp\left(-\frac{1}{2\hbar_{\nu}}\sum_{i,j=1}^n\bar{\nu}_i(t)\delta_{ij}(t)\bar{\nu}_j(t)\right),\\ \nonumber
	&=\exp\left(-\frac{\epsilon \hbar_{\nu}}{2}\sum_{i,j=1}^np_i^2\right).
\end{align}
Inverting the Fourier transform, it follows that
\begin{align}\label{eq:InfPropFT0}  
	P(t+\epsilon,x|t,x')&\approx \frac{1}{(2\pi)^n}\int\left\{ d^np \right\}\exp\left(i\sum_{i=1}^np_i(x_i-x_i'-\epsilon f_i(x'))-\epsilon\frac{\hbar_{\nu}}{2}\sum_{i=1}^np_i^2\right),\\ \nonumber
	&=\frac{1}{\sqrt{(2\pi\hbar_{\nu}\epsilon)^n}}\exp\left[ -\frac{1}{2\hbar_{\nu}\epsilon}\sum_{i=1}^n\left( x_i-x_i'-\epsilon f_i(x') \right)^2 \right].
\end{align}
Partition the time interval $[t_0,t]$ into $N$ equi-spaced time intervals so that $t_i=t_0+i\epsilon$ where $\epsilon=(t-t_0)/N$. Then, from the Chapman-Kolmogorov semi-group property it follows that
\begin{align}\label{eq:SplitandCK}
	P(t,x|t_0,x_0)&=\int\{d^nx(t_1)\cdots d^nx(t_{N-1})\}P(t,x|t_{N-1},x(t_{N-1}))\cdots P(t_1,x(t_1)|t_0,x_0),\\ \nonumber
	&=\int\left\{ \prod_{i=1}^{N-1}d^nx(t_i) \right\}\left[\prod_{i=1}^{N}P(t_i,x(t_i)|t_{i-1},x(t_{i-1}))\right],
\end{align}
where $x(t_0)=x_0$ and $x(t_N)=x$. 
Therefore, 
\begin{align}
	P(t,x|t_0x_0)=\int_{x(t_0)=x_0}^{x(t)=x}\left[ \mc{D}x(t) \right]\exp\left[ -\frac{1}{\hbar_{\nu}}S_{I}(t_0,t) \right],
\end{align}
where
\begin{align}
	S_{I}(t_0,t)&=\lim_{\epsilon\ra0}\frac{1}{2\epsilon}\sum_{k=1}^N\sum_{i=1}^n\left( x_i(t_k)-x_i(t_{k-1})-\epsilon f(x_i(t_{k-1})) \right)^2,\\ \nonumber
	[\mc{D}x(t)]&=\lim_{\epsilon\ra0}\frac{1}{\sqrt{(2\pi\hbar_{\nu}\epsilon)^n}}\prod_{i=1}^{N-1}\left\{ \frac{d^nx(t_i)}{\sqrt{(2\pi\hbar_{\nu}\epsilon)^n}} \right\},\qquad\epsilon=\frac{t-t_0}{N}.
\end{align}
Observe that the extra factor of $1/\sqrt{(2\pi\hbar_{\nu}\epsilon)^n}$ is due to $N$ terms in the square bracketed term in Equation \ref{eq:SplitandCK}. In the continuum notation, this is
\begin{align}\label{eq:PIformulaItoEasy}
	P(t,x|t_0,x_0)=\int_{x(t_0)=x_0}^{x(t)=x}\left[ \mc{D}x(t) \right]\exp\left[ -\frac{1}{2}\int_{t_0}^td\tau\sum_{i=1}^n(\dot{x}_i-f_i(x^{(0)}(t)))^2 \right],
\end{align}
where the superscript $(0)$ is to remind us that the drift is evaluated at the earlier time. This is termed the ``pre-point'' version.

This formula clearly differs from the formula (Equation \ref{eq:PIformula}) obtained using functional methods. In fact, another important difference is that in the Equation \ref{eq:PIformulaItoEasy}, the argument of $f$ is to be evaluated at the pre-point, rather than the mid-point (as in the case of Equation \ref{eq:PIformula}. This discrepancy is resolved once the pre-point form of the PI formula is converted to the symmetric form.  

In order to obtain the result in the Feynman convention, the argument of $f$ needs to be symmetrized. It is straightforward to obtain a more general expression as follows. As evident from Equations \ref{eq:DiscLangSimple} and \ref{eq:DiscNoise00}, typical values of $x-x'\sim O(\epsilon^{1/2})$. Let $x^{(r)}=x'+r(x-x')$. Then, it follows that the leading order term in $x_i-x_i'-\epsilon f_i(x')$ is given by
\begin{align}
	x_i-x_i'-\epsilon f_i(x')=x_i-x_i'-\epsilon f_i(x^{(r)})+r\epsilon\sum_{i'=1}^n(x_{i'}-x_{i'}')\frac{\partial f_i}{\partial x_{i'}}(x^{(r)})+O(\epsilon^2)(\bar{x}).
\end{align}
The last term may be eliminated by introducing a new variable $\tilde{x}_i$ defined as
\begin{align}\label{eq:Transf00}
	\tilde{x}_i=x_i+r\epsilon \sum_{i'=1}^n(x_{i'}-x_{i'}')\frac{\partial f_i}{\partial x_{i'}}(x^{(r)}).
\end{align}
The Jacobian under this transformation is ($d^n\tilde{x}=\det(\partial \tilde{x}_i)/\partial x_i)d^nx$)
\begin{align}
	\det\left[ \frac{\partial}{\partial x_i}\left( x_j+r\epsilon(x_l-x_l')\frac{\partial f_j}{\partial x_l}(x^{(r)}) \right) \right]^{-1}=\exp\left[ -\frac{1}{2}\epsilon \sum_{i=1}^n\frac{\partial f_i}{\partial x_i}(x^{(r)})\right],
\end{align}
where the identity $\det A=\exp(\text{tr}\ln A)$ has been used. Note that the transformation in Equation \ref{eq:Transf00}  does not change the $x-x'$ term in $S_{\epsilon}(t_0,t)$  since the difference is $O(\epsilon^{3/2})$. Thus the overall exponent is
\begin{align}
	S_{\epsilon}(t_0,t)=\lim_{\epsilon\ra0}\sum_{k=1}^N\sum_{i=1}^n\left[\frac{1}{2\epsilon}\left( x_i(t_k)-x_i(t_{k-1})-\epsilon f_i(\bar{x}(t_k)) \right)^2+\frac{1}{2}\epsilon\frac{\partial f_i}{\partial x_i}(\bar{x}(t_k))\right]. 
\end{align}

Therefore, the result for the transition probability density  is
\begin{align}
	P(t,x|t_0,x_0)=\int_{x(t_0)=x_0}^{x(t)=x}\left[ \mc{D}x(t) \right]\exp\left( -\frac{1}{\hbar_{\nu}}S(t_0,t) \right),
\end{align}
where 
\begin{align}
	S(t_0,t)=\sum_{i=1}^n\int_{t_0}^t\left[ \frac{1}{2}(\dot{x}_i(t)-f_i(x^{(r)}(t)))^2+\hbar_{\nu}r\frac{\partial f_i}{\partial x_i}(x^{(r)}(t)) \right]d\tau.
\end{align}
When $r=\frac{1}{2}$, or the mid-point or Feynman discretization, agreement with the result obtained by functional methods is seen. The $r=0$ case is, of course, the pre-point version. The exponent, $S(t,t_0)$, is referred to as the action in physics literature.

\section{Path Integral Formula II: Explicit Time Dependence}\label{sec:pathinteg2}

In this section, the general additive noise model is considered: noise is colored and the diffusion vielbein is a time-dependent rectangular matrix. Since the diffusion vielbein is no longer invertible, the method used in the previous section is not applicable. 

The state model is given by
\begin{align}
\label{eq:LangMult01}
d\rv{x}(t)=f(\rv{x}(t),t)dt+e(t)d\rv{\nu}(t),  
\end{align}
where $\rv{x}(t),f(\rv{x}(t))\in\mathbb{R}^n,\rv{\nu}(t)\in\mathbb{R}^p$, and  $e(t)\in\mathbb{R}^{n\times p}$. Observe that drift is now an explicit function of time and the diffusion vielbein, apart from being a rectangular matrix, has an explicit time dependence as well. However, the diffusion term is assumed to be independent of the state vector; the more general state-dependent, or multiplicative noise, case is more subtle and will be studied in a future paper. Again, the It\^o and Stratanovich forms are the same for this equation. A sample of this stochastic process satisfies the following differential equation 
\begin{align}
	\dot{x}_i(t)=f_i(x(t),t)+\sum_{a=1}^pe_{ia}(t)\nu_a(t),\quad i=1,2,\dots,n,
\end{align}
where the $\nu$ is a sample of a Gaussian noise process with $\delta-$correlated covariance matrix $Q$. 

It is assumed that the diffusion matrix is invertible; if not, it is physically plausible that results for the singular case are close to the ``nearly singular'' case, which is obtained by adding a small quantitity to render it invertible.
\subsection{Formal Derivation Using Functional Calculus}

The discussion here is similar to that in Section \ref{sec:FormalPlain}. However, imposing a delta function constraint requires a different method (as in Section \ref{sec:DeltaFnConstraints}) since $e(t)$ is not a square matrix; and hence is not invertible. The path integral expression for the transition probability density is
\begin{align}
	\vev{\delta^n[\rv{x}(t)-x]}|_{x(t_0)=x_0}&=\int [\mc{D}\rho(\nu)]\delta^n\left[ x(t)-x \right]|_{x(t_0)=x_0}, \\ \nonumber
	&=\int\left[ \mc{D}\nu(t) \right]\exp\left( -\frac{1}{2}\sum_{a,b=1}^p\int \nu_a(t)(Q^{-1}(t))_{ab}\nu_b(t)dt \right)\times\delta[x(t)-x]|_{x(t_0)=x_0},
\end{align}
where $x(t)$ satisfies the Langevin equation
\begin{align}
\label{eq:LangConstraint}
\dot{x}_i(t)=f_i(x(t),t)+\sum_{a=1}^pe_{ia}(t)\nu_a(t).	
\end{align}
The Langevin equation condition can be imposed Equation \ref{eq:UnityFunctional} to yield 
\begin{align}
	P(t,x|t_0,x_0)&=\int\left[ \mc{D}\nu(t) \right]\exp\left( -\frac{1}{2}\sum_{a,b=1}^p\int \nu_a(t)(Q^{-1}(t))_{ab}\nu_b(t)dt \right)\\ \nonumber
	&\quad\times\int\left[ \mc{D}\dot{x}(t') \right]J\delta^n[\dot{x}_i(t)-f_i(x(t),t)-\sum_{a=1}^pe_{ia}\nu_a(t)], \\ \nonumber
	&\qquad\times\delta[x(t)-x]|_{x(t_0)=x_0}.
\end{align}
Upon evaluating the functional derivative of the Langevin equation
\begin{align}
	\frac{\delta}{\delta x_j(t')}\left[ \dot{x}_i(t)-f_i(x(t),t)-\sum_{a=1}^pe_{ia}(t)\nu_a(t) \right]&=\left[ \delta_{ij}\frac{d}{dt}-\frac{\partial f_i}{\partial x_j}(x(t'),t') \right]\delta(t-t'),\\ \nonumber
	&=\frac{d}{dt}\left[ \delta_{ij}\delta(t-t')-\theta(t-t')\frac{\partial f_i}{\partial x_j}(x(t'),t') \right],
\end{align}
it follows that the Jacobian $J$ is given by 
\begin{align}
	J&=\det\left( \frac{d}{dt} \right)\det\left( \delta_{ij}\delta(t-t')-\theta(t-t')\frac{\partial f_i}{\partial x_j}(x(t'),t') \right),\\ \nonumber
	&=\mc{N}\det\left( \delta_{ij}\delta(t-t')-\theta(t-t')\frac{\partial f_i}{\partial x_j}(x(t'),t')) \right),
\end{align}
where $\mc{N}$ is an irrelevant constant that can be ignored (or absorbed into the measure). Hence, 
\begin{align}
	\ln J&=\ln\det\left[ \delta_{ij}\delta(t-t')-\theta(t-t')\frac{\partial f_i}{\partial x_j}(x(t'),t') \right],\\ \nonumber
	&=-\sum_{i=1}^n\frac{1}{2}\int\frac{\partial f_i}{\partial x_i}(x(t),t)dt.
\end{align}
Also, following arguments in Equation \ref{eq:Measxdot} the measure $[\mc{D}\dot{x}(t)]$ can be replaced by $[\mc{D}x(t)]$. Thus, so far, 
\begin{align}
	P(t,x|t_0,x_0)=&\int_{x(t_0)=x_0}^{x(t)=x}\left[ \mc{D}x(t) \right]\left[ \mc{D}\nu(t) \right]\exp\left( -\frac{1}{2}\int\sum_{i,j=1}^n\sum_{a,b=1}^p\nu_a(t)(Q^{-1}(t))_{ab}\nu_b(t)dt \right)\\ \nonumber
	&\quad\times\delta^n\left(\dot{x}_i(t)-f_i(x(t),t)-\sum_{a=1}^pe_{ia}(t)\nu_a(t)\right)\exp\left( -\frac{1}{2}\int\sum_{i=1}^n\frac{\partial f_i}{\partial x_i}(x(t),t)dt \right).
\end{align}
Using the Fourier integral version of the delta function, the transition probability density becomes
\begin{align}
	P(t,x|t_0,x_0)=&\int_{x(t_0)=x_0}^{x(t)=x}\left[ \mc{D}x(t) \right]\left[ \mc{D}\nu(t) \right]\left[ \mc{D}\lambda(t) \right]\\ \nonumber
	&\qquad\times\exp\left( i\int\sum_{i=1}^n\lambda_i(t)(\dot{x}_i(t)-f_i(x(t),t)-\sum_{a=1}^pe_{ia}(t)\nu_a(t)) \right) \\ \nonumber
	&\times\exp\left( -\frac{1}{2}\int\sum_{a,b=1}^p\nu_a(t)(Q^{-1}(t))_{ab}\nu_b(t)dt \right)\times\exp\left( -\frac{1}{2}\int\sum_{i=1}^n\frac{\partial f_i}{\partial x_i}(x(t),t)dt \right).
\end{align}
Integration over $\nu(t)$  results in 
\begin{align}
	P(t,x|t_0,x_0)=&\int_{x(t_0)=x_0}^{x(t)=x}\left[ \mc{D}x(t) \right]\left[ \mc{D}\lambda(t) \right] \exp\left(-\int dt\left[\frac{1}{2}\sum_{i=1}^n\frac{\partial f_i}{\partial x_i}(x(t),t)\right] \right)\\ \nonumber
	&\times\exp\left( -\int dt\left[ \sum_{i,j=1}^n\lambda_i(t)g_{ij}(t)\lambda_j(t)-i\sum_{i-1}^n\lambda_i(t)\left( \dot{x}_i(t)-f_i(x(t),t) \right)\right]\right).	
\end{align}
Finally, integrating over $\lambda_i(t)$ leads to 
\begin{align}\label{eq:TransProbTimeDepFinal}
	P(t,x|t_0,x_0)=\int_{x(t_0)=x_0}^{x(t)=x}[\mc{D}x(t)]\exp(-S),
\end{align}
where the ``action'' $S$ is  
\begin{align}
	S= \frac{1}{2}\int dt\left[ \sum_{i,j=1}^n(\dot{x}_i(t)-f_i(x(t),t))g_{ij}^{-1}(t)\left( \dot{x}_j(t)-f_j(x(t),t) \right) +\sum_{i=1}^n \frac{\partial f_i}{\partial x_i}(x(t),t)\right].
\end{align}

\subsection{Derivation Using the Chapman-Kolmogorov Semigroup Property}

Integrating the Equation \ref{eq:LangMult01} over an infinitesimal time interval $(t,t+\epsilon)$ yields
\begin{align}
	\label{eq:LangEqMultInfinitesimal}
	\rv{x}_i(t+\epsilon)=\rv{x}_i(t)+\epsilon f_i(\rv{x}(t),t)+\sum_{a=1}^p\int_t^{t+\epsilon}e_{ia}(t)\rv{\nu}_a(t)+O(\epsilon^{3/2}).
\end{align}
Note, in particular, that replacing $t$ with $t+\alpha\epsilon,$ in $e(t)$ and $f(\rv{x}(t),t)$(where $\alpha\in(0,1)$) leads to errors of order higher than $\epsilon$ $(O(\epsilon^{3/2}))$ and hence are irrelevant in the continuum limit. The continuum Langevin equation (Equation \ref{eq:LangMult01}) then becomes the following discretized Langevin equation:
\begin{align}
\label{eq:LangEqMulDis}
\rv{x}_i(t+\epsilon)=\rv{x}_i(t)+\epsilon f_i(\rv{x}(t),t)+\sqrt{\epsilon}\sum_{a=1}^pe_{ia}\bar{\rv{\nu}}_a(t).
\end{align}
As in the previous discussion, the following identification has been made
\begin{align}
	\label{eq:LangEqMulDis01}
	\sqrt{\epsilon}\sum_{a=1}^pe_{ia}(t)\bar{\rv{\nu}}_a(t)\equiv\sum_{a=1}^p\int_t^{t+\epsilon} e_{ia}(\tau)\rv{\nu}_a(\tau)d\tau.
\end{align}
Since 
\begin{align}
	\vev{\rv{\nu}_a(t)}&=0,\\ \nonumber
	\vev{\rv{\nu}_a(t)\rv{\nu}_b(t')}&=Q_{ab}(t)\delta(t-t'),
\end{align}
it is clear that
\begin{align}
	\epsilon\sum_{a,b=1}^p\vev{e_{ia}(t)\bar{\rv{\nu}}_a(t)\bar{\rv{\nu}}_b(t')e^T_{bj}(t')}&=\epsilon \sum_{a,b=1}^pe_{ia}(t)\vev{\bar{\rv{\nu}}_a(t)\bar{\rv{\nu}}_b(t')}e^T_{bj}(t),\\ \nonumber
	&=\int_t^{t+\epsilon}\int_{t'}^{t'+\epsilon}\sum_{a,b=1}^pe_{ia}(\tau)\vev{\rv{\nu}_a(\tau)\rv{\nu}_b(\tau')}e^T_{bj}(\tau')d\tau d\tau',\\ \nonumber
	&=\int_t^{t+\epsilon}\int_{t'}^{t'+\epsilon}\sum_{a,b=1}^pe_{ia}(\tau)Q_{ab}(\tau)\delta(\tau-\tau')e_{bj}^T(\tau')d\tau d\tau',\\ \nonumber
	&=
	\begin{cases}
		\int_t^{t+\epsilon}\sum_{a,b=1}^pe_{ia}(\tau)Q_{ab}(\tau)e_{bj}^Td\tau,&t=t',\\
		0,&t\neq t'
	\end{cases},\\ \nonumber
	&=
	\begin{cases}
		\epsilon \sum_{a,b=1}^pe_{ia}(\tau)Q_{ab}(\tau)e_{bj}^T(\tau), &t=t',\\
		0,&t\neq t'
	\end{cases}.
\end{align}
Hence, it follows that
\begin{align}
	\label{eq:NoiseMultDisc}
	\vev{\bar{\rv{\nu}}_a(t)}&=0,\\ \nonumber
	\vev{\bar{\rv{\nu}}_a(t)\bar{\rv{\nu}}_b(t')}&=Q_{ab}(t)\delta_{tt'},
\end{align}
and $\bar{\rv{\nu}}(t)$ is a discrete-time Gaussian process. 

The Fourier transform of $P(t+\epsilon,x;t,x')$ with respect to $x$ is 
\begin{align} 
	\label{eq:FourierPMult00A}
	\tilde{P}(t+\epsilon,p|t,x')&=\int d^nx\exp\left(-i\sum_{i=1}^np_ix_i\right)P(t+\epsilon,x;t,x').
\end{align}
Interchanging the order of averaging and integration over $x$ leads to 
\begin{align}\label{eq:FourierPMult00B}
\tilde{P}(t+\epsilon,p|t,x')&=\vev{\int d^nx\exp\left(-i\sum_{i=1}^np_ix_i\right)\delta^n(\rv{x}(t+\epsilon)-x)},\\ \nonumber
	&=\vev{\exp\left(-i\sum_{i=1}^np_i\rv{x}_i(t+\epsilon)\right)}.
\end{align}
Substituting the value of $\rv{x}(t+\epsilon)$ in Equation \ref{eq:LangEqMulDis}, 
\begin{align}\label{eq:FourierPMult01}
	\tilde{P}(t+\epsilon,p|t,x')&=\vev{\exp\left(-i\sum_{i=1}^np_i\rv{x}_i(t+\epsilon)\right)},\\ \nonumber
	&=\exp\left(-i\sum_{i=1}^np_i(x_i'+\epsilon f_i(x',t))\right)\vev{\exp\left(-i\sqrt{\epsilon}\sum_{i=1}^n\sum_{a=1}^pp_ie_{ia}(t)\bar{\rv{\nu}}_a(t)\right)}.
\end{align}
Now,
\begin{align}\label{eq:AvexpPNu}
	&\vev{\exp\left(-i\sqrt{\epsilon}\sum_{i=1}^n\sum_{a=1}^pp_ie_{ia}(t)\bar{\rv{\nu}}_{a}(t)\right)}=\\ \nonumber
	&\quad\int \frac{1}{\sqrt{(2\pi\epsilon)^n\det Q(t)}}\left\{ d^n\bar{\nu}(t) \right\}\exp\left(-i\sqrt{\epsilon}\sum_{i=1}^n\sum_{a=1}^pp_ie_{ia}\bar{\nu}_a(t)-\frac{\epsilon}{2}\sum_{a,b=1}^p\bar{\nu}_a(t)Q_{ab}^{-1}(t)\bar{\nu}_b(t)\right),\\ \nonumber
	&\quad=\exp\left(-\frac{\epsilon}{2}\sum_{i,j=1}^n\sum_{a,b=1}^pp_ie_{ia}(t)Q_{ab}(t)e^T_{bj}(t)p_j\right).
\end{align}
Combining Equations \ref{eq:AvexpPNu} and \ref{eq:FourierPMult01}, it is evident that 
\begin{align}
	\label{eq:FourierPMult02}
	\tilde{P}(t+\epsilon,p|t,x')&=\exp\left(-\frac{\epsilon}{2}\sum_{i,j=1}^n\sum_{a,b=1}^pp_ie_{ia}(t)Q_{ab}(t)e^T_{bj}p_j-i\sum_{i=1}^np_i(x_i'+\epsilon f_i(x',t))\right).
\end{align}
Inverting the Fourier transform, 
\begin{align} \label{eq:InfPropFT} 
	P(t+\epsilon,x|t,x')&=\frac{1}{(2\pi)^n}\int \left\{d^np\right\}\exp\left(i\sum_{i=1}^np_ix_i\right)\tilde{P}(t+\epsilon,p; t,x'),\\ \nonumber
	&=\frac{1}{(2\pi)^n}\int \left\{d^np\right\}\exp\left(-\frac{\epsilon}{2}\sum_{i,j=1}^n\sum_{a,b=1}^pp_ie_{ia}(t)Q_{ab}(t)e_{bj}^T(t)p_j\right)\\ \nonumber
	&\quad\times\exp\left(+i\sum_{i=1}^np_i(x_i-x_i'-\epsilon f_i(x',t))\right),\\ \nonumber
	&=\frac{1}{\sqrt{(2\pi\epsilon)^n\det e(t)Q(t)e^T(t)}}\exp\left( -S_{\epsilon}(t+\epsilon,t)\right),
\end{align} 
where
\begin{align}
	\label{eq:ActionMultDisc}
	S_{\epsilon}(t+\epsilon,t)=+\frac{1}{2\epsilon}\sum_{i,j=1}^n\left[(x_i-x_i'-\epsilon f_i(x',t))\left[\sum_{a,b=1}^pe_{ia}(t)Q_{ab}(t)e_{bj}^T(t) \right]^{-1}(x_j-x'_j-\epsilon f_j(x',t))\right].
\end{align}
From this result, the finite time result can be calculated using the Chapman-Kolmogorov semi-group property of the transition probability density. Specifically, divide the time interval $(t,t_0)$ into $N$ parts,  i.e., $\epsilon=(t-t_0)/N$, and use Equation \ref{eq:SplitandCK} to get
\begin{align}
	P(t,x|t_0,x_0)=\lim_{\epsilon\ra0}\int\frac{1}{\sqrt{(2\pi\epsilon)^n\det g(t_0)}}\prod_{i=1}^{N-1}\left\{\frac{d^nx(t_0+i\epsilon)}{\sqrt{(2\pi\epsilon)^n\det g(t_0+i\epsilon)}}  \right\}\exp{-S_{\epsilon}(t_0,t)},
\end{align}
where 
\begin{align}
	S_{I}(t,t_0)=\lim_{\epsilon\ra0}\frac{1}{2\epsilon}\sum_{k=1}^N\sum_{i,j=1}^n &\Big[ \left( x_i(t_k)-x_i(t_{k-1})-\epsilon f_i(x(t_{k-1}),t_{k-1}) \right)\\ \nonumber
	&\qquad\times g^{-1}_{ij}(t_k)\left( x_j(t_k)-x_j(t_{k-1})-\epsilon f_j(x(t_{k-1}),t_{k-1}) \right) \Big],
\end{align}
and 
\begin{align}
	g_{ij}(t)=\left( \sum_{a,b=1}^pe_{ia}(t)Q_{ab}(t)e^T_{bj}(t) \right).
\end{align}
When taking the continuum limit, the argument of $f$ needs to be symmetrized in the Feynman form. Note that it suffices to symmetrize $x$ in the argument since $x-x'\approx O(\sqrt{\epsilon})$. Symmetrization  with respect to $t$, i.e., $t_{k-1}\ra \frac{1}{2}(t_k+t_{k-1})$, leads to an error higher than $O(\epsilon)$ and vanishes in the continuum limit.

Thus, setting $\bar{x}(t_k)=\frac{1}{2}(x(t_k)+x(t_{k-1}))$,
\begin{align}
	x_i(t_k)-x_i(t_{k-1})-\epsilon f_i(x(t_{k-1}), t_k)&=x_i(t_k)-x_i(t_{k-1})-\epsilon f_i\left( \bar{x}(t_k)-\frac{x(t_k)-x(t_{k-1})}{2},t_k \right),\\ \nonumber
	&=x_i(t_{k})-x_i(t_{k-1})+\frac{\epsilon}{2}\sum_{j=1}^n(x_j(t_k)-x_j(t_{k-1}))\frac{\partial f_i}{\partial \bar{x}_j}(\bar{x},t).
\end{align}
Define $\tilde{x}$ as follows: 
\begin{align}
	\tilde{x}_i=x_i+\frac{\epsilon}{2}\sum_{j=1}^n\left( x-x' \right)_j\frac{\partial f_i}{\partial x_j}(\bar{x},t).
\end{align}
The Jacobian under this change of variables (from $x$ to $\tilde{x}$) is 
\begin{align}\label{eq:PIFormulaTD}
	\det\left[ \frac{\partial}{\partial x_i}\left( x_j+\frac{\epsilon}{2}\sum_{l=1}^n(x-x')_l\frac{\partial f_j}{\partial x_l} \right)  \right]^{-1}&=\det\left[ \delta_{ij}+\frac{\epsilon}{2}\frac{\partial f_j}{\partial x_i} \right]^{-1},\\ \nonumber
	&=\exp\left( -\frac{\epsilon}{2}\sum_{i=1}^n\frac{\partial f_i}{\partial x_i} \right),
\end{align}
where the identity $\det(1+\epsilon A)\approx \exp \epsilon\text{tr} A$ has been used. Hence, the overall action is 
\begin{align}\label{eq:PIformulaGeneral00}
	&\frac{1}{2\epsilon}\sum_{k=1}^N\sum_{i,j=1}^n\left[\left( x_{i}(t_k)-x_{i}(t_{k-1})-\epsilon f_i(\bar{x}_{i,k},t_k) \right)g^{-1}_{ij}(t_k)\left( x_j(t_k)-x_j(t_{k-1})+\epsilon f_j(\bar{x}_{k},t_{k}) \right)+\frac{\epsilon}{2}\frac{\partial f_i}{\partial x_i}(\bar{x}(t_k),t_k)\right].
\end{align}

Therefore, in the continuum limit, the transition probability density is given by
\begin{align}
	\label{eq:TransProbTimeDep00}
	P(t,x|t_0,x_0)=\int_{x(t_0)=x_0}^{x(t)=x}\left[ \mc{D}x(t) \right]\exp(-S_F(t_0,t)),
\end{align}
where the action is 
\begin{align}
	\label{eq:ActionProbTimeDep00}
	S_F(t,t_0)=\frac{1}{2}\sum_{i=1}^n\int_{t_0}^t\left[ [\dot{x}_i(t)-f_i(x(t),t)]g_{ij}^{-1}(t)[\dot{x}_j(t)-f_j(x(t),t)]+\frac{\partial f_i}{\partial x_i}(x(t),t) \right],
\end{align}
and
\begin{align}
	[\mc{D}x(t)]=\lim_{\epsilon\ra0}\frac{1}{\sqrt{(2\pi\epsilon)^n\det g(t_0)}}\prod_{i=1}^{N-1}\left\{d^nx(t_0+i\epsilon)\frac{1}{\sqrt{(2\pi\epsilon)^n\det g(t_0+i\epsilon)}}  \right\}.
\end{align}

It is clear that for the general discretization the action is
\begin{align}\label{eq:InfPropFT1}
	S^{(r)}(t,t_0)=\sum_{i=1}^n\int_{t_0}^t\left[ \frac{1}{2}\left( \dot{x}_i(t)-f_i(x^{(r)}(t),t) \right)g_{ij}^{-1}(t)\left( \dot{x}_j(t)-f_j(x^{(r)}(t),t)\right)+r\frac{\partial f_i}{\partial x_i}(x^{(r)}(t),t) \right].
\end{align}

Na\i vely, it seems that the path integral formula depends on $r$, which contradicts the fact that the FPKfe and the Langevin equation do not depend on $r$. This is resolved in Section\ref{sec:Verification} where it is demonstrated that the $r$ dependence of the Feynman path integral formula is illusory, i.e., $r$ dependence cancels.

\section{Verification of the Path Integral Formula}\label{sec:Verification}

In this section, it is independently verified that the path integral formulas derived in the previous sections are the fundamental solutions of the FPKfe. Clearly, it is sufficient to prove the path integral formula derived for the explicit time-dependent case and for arbitrary discretization. The method used closely follows the techniques first used by Feynman to verify the path integral formula he derived for the fundamental solution of the Schr\"odinger equation (see \cite{R.P.FeynmanandA.R.Hibbs1965}).

From the Chapman-Kolmogorov semi-group property, it follows that 
\begin{align}
	P(t+\epsilon,x|t,x_0)=\int P(t+\epsilon,x|t,x')P(t,x'|t_0,x_0)\{d^nx'\}.
\end{align}
According to Equation \ref{eq:InfPropFT} and \ref{eq:InfPropFT1} 
\begin{align}\label{eq:CKVerify}
	P&(t+\epsilon,x|t,x')=	\frac{1}{\sqrt{(2\pi\epsilon)^n\det g(t)}} \\ \nonumber
	&\quad\times\exp\left(-\frac{\epsilon}{2}\sum_{i,j=1}^n\left[ \frac{x_i-x_i'}{\epsilon}-f_i(x^{(r)},t) \right]g^{-1}_{ij}(t)\left[ \frac{x_j-x_j'}{\epsilon}-f_j(x^{(r)},t) \right]-\epsilon\frac{1}{2}\sum_{i=1}^n\frac{\partial f_i}{\partial x_i}(x^{(r)},t)\right).
\end{align}
Here $x^{(r)}=x'+r(x-x')$. 

The quantity $P(t+\epsilon,x|t,x')$ is large only if 
\begin{align}
	\frac{x-x'}{\epsilon}\approx f(x^{(r)},t).
\end{align}
We may then write
\begin{align}
	x=x'+\epsilon f(x^{(r)},t)+\eta,
\end{align}
or
\begin{align}
	x'=x-\epsilon f(x^{(r)},t)-\eta,
\end{align}
where the equalities are valid to $O(\epsilon)$. Substituting this into Equation \ref{eq:CKVerify}, 
\begin{align}\label{eq:FeynVerify00A}
	P(t+\epsilon,x|t_0,x_0)&=\int_{-\infty}^{\infty}\exp\left(-\frac{1}{2\epsilon}\sum_{i,j=1}^n\eta_ig^{-1}_{ij}(t)\eta_j\right) \\ \nonumber
	& \quad\times\left( 1-r\epsilon\sum_{i=1}^n\frac{\partial f_i}{\partial x_i} (\bar{x},t)\right)P(t,x'|t_0,x_0)\left\{ d^nx' \right\}\frac{1}{\sqrt{(2\pi\epsilon)^n\det g}}.\\ 
\end{align}
Next, the variable of integration is changed from $x'$ to $\eta$. Since
\begin{align}
	\eta=x'-x+\epsilon f(x'+r(x-x').t),
\end{align}
so  that
\begin{align}
	\frac{\partial\eta_i}{\partial x'_j}=\delta_{ij}+(1-r)\epsilon\frac{\partial f_i}{\partial x'_j}(x^{(r)},t),	
\end{align}
and the Jacobian of the transformation from $x'$ to $\eta$ is
\begin{align}
	\left( 1-(1-r)\epsilon\sum_{i=1}^n\frac{\partial f_i}{\partial x_i} \right).
\end{align}
Combining these results leads to
\begin{align}\label{eq:FeynVerify00B}
P(t+\epsilon,x|t_0,x_0)&=\int_{-\infty}^{\infty}\exp\left(-\frac{1}{2\epsilon}\sum_{i,j=1}^n\eta_ig^{-1}_{ij}(t)\eta_j\right)& \\ \nonumber
& \quad\times\left( 1-\frac{\epsilon}{2}\sum_{i=1}^n\frac{\partial f_i}{\partial x_i} \right)P(t,x'|t_0,x_0)\left( 1-(1-r)\epsilon\sum_{i=1}^n\frac{\partial f_i}{\partial x_i} \right)\left\{d^n\eta\right\}\frac{1}{\sqrt{(2\pi\epsilon)^n\det g}}, \\ \nonumber
&=\int_{-\infty}^{\infty}\exp\left(-\frac{1}{2\epsilon}\sum_{i,j=1}^n\eta_ig^{-1}_{ij}(t)\eta_j\right)& \\ \nonumber
& \quad\times\left( 1-\epsilon\sum_{i=1}^n\frac{\partial f_i}{\partial x_i} \right)P(t,x-\epsilon f(x,t)-\eta|t_0,x_0)\left\{ d^n\eta \right\}\frac{1}{\sqrt{(2\pi\epsilon)^n\det g}},
\end{align}
where the Jacobian of the the transformation from $x'$ to $\eta$ to $O(\epsilon)$ is included.

The left hand side of Equation \ref{eq:FeynVerify00B} is
\begin{align}\label{eq:LHSVer}
	P(t,x|t_0,x_0)+\epsilon\frac{\partial P}{\partial t}(t,x|t_0,x_0).
\end{align}
Also,
\begin{align}
	P(t,x-\epsilon f(x,t)-\eta|t_0,x_0)=P(t,x|t_0,x_0)-\epsilon(f_i(x,t)+\eta_i)\frac{\partial P}{\partial x_i}(t,x|t_0,x_0)+\frac{1}{2}\eta_i\eta_j\frac{\partial^2P}{\partial x_i\partial x_j}(t,x|t_0,x_0),
\end{align}
where terms that contribute to the order of $\epsilon$ have been retained. Therefore, the integrand 
\begin{align}
	\left( 1-\epsilon\sum_{i=1}^n\frac{\partial f_i}{\partial x_i}(x,t) \right)\left( P(t,x|t_0,x_0)-\epsilon(f_i(x,t)+\eta_i)\frac{\partial P}{\partial x_i}(t,x|t_0,x_0)+\frac{1}{2}\eta_i\eta_j\frac{\partial^2P}{\partial x_i\partial x_j}(t,x|t_0,x_0 \right)
\end{align}
becomes, to $O(\epsilon)$,
\begin{align}
	P(t,x|t_0,x_0)-\epsilon\sum_{i=1}^n\frac{\partial}{\partial x_i}\left[ f_i(x,t)P(t,x|t_0,x_0) \right]+\frac{1}{2}\eta_i\eta_j\frac{\partial^2P}{\partial x_i\partial x_j}(t,x|t_0,x_0).
\end{align}
Here, the term linear in $\eta$ has been dropped since its contribution to $P(t+\epsilon,x|t_0,x_0)$ vanishes upon integration over $\eta$ due to odd symmetry.

Since
\begin{align}
	\int_{-\infty}^{\infty}\exp\left( -\frac{1}{2\epsilon}\sum_{i,j=1}^n\eta_ig^{-1}_{ij}(t)\eta_j \right)\left\{ d^n\eta \right\}\frac{1}{\sqrt{(2\pi\epsilon)^n\det g(t)}}&=1,\\ \nonumber
	\int_{-\infty}^{\infty}\eta_i\eta_j\exp\left( -\frac{1}{2\epsilon}\sum_{i',j'=1}^n\eta_{i'}g^{-1}_{i'j'}(t)\eta_{j'} \right)\left\{ d^n\eta \right\}\frac{1}{\sqrt{(2\pi\epsilon)^n\det g(t)}}&=\epsilon g_{ij}(t),
\end{align}
the right hand side of Equation \ref{eq:FeynVerify00B} is
\begin{align}\label{eq:RHSVer}
	P(t,x|t_0,x_0)-\epsilon\sum_{i=1}^n\frac{\partial}{\partial x_i}\left[ f_i(x,t)P(t,x|t_0,x_0) \right]+\epsilon\frac{1}{2}\sum_{i,j=1}^ng_{ij}(t)\frac{\partial^2P}{\partial x_i\partial x_j}(t,x|t_0,x_0).
\end{align}
Finally, note that 
\begin{align}\label{eq:DeltaFnVerify}
	P(t,x|t,x')&=\lim_{\epsilon\ra0}\frac{1}{\sqrt{(2\pi\epsilon)^n\det g(t)}}\exp\left( -\frac{1}{2\epsilon}\sum_{i,j=1}^n\left[ (x_i-x_i')g^{-1}_{ij}(t)(x_j-x_j') \right] \right),\\ \nonumber
	&=\delta^n(x-x').
\end{align}

Therefore, from Equations \ref{eq:LHSVer}, \ref{eq:RHSVer}, and \ref{eq:DeltaFnVerify} it follows that $P(t,x|t_0,x_0)$ is the fundamental solution of the FPKfe:
\begin{align}
	{\left\lbrace\begin{aligned}	
		\frac{\partial P}{\partial t}(t,x|t_0,x_0)&=-\sum_{i=1}^n\frac{\partial}{\partial x_i}\left[ f_i(x,t)P(t,x|t_0,x_0) \right]+\frac{1}{2}\sum_{i,j=1}^n\frac{\partial^2}{\partial x_i\partial x_j}\left[g_{ij}(t)P(t,x|t_0,x_0))\right], \\ 
	P(t,x|t,x_0)&=\delta^n(x-x_0).\end{aligned}
\right.}
\end{align}

\section{Examples}

In this section, a few examples are presented illustrating the utility of the path integral formulas derived in this paper. For simplicity, all examples are one-dimensional. Note that an accurate determination of the transition probability density automatically implies that the continuous-filtering problem can be solved very accurately. It is therefore sufficient to demonstrate that the transition probability density can be computed accurately. 

From the discussions in the previous sections, the one-step  $r=0$, or `pre-point', approximate formula is
\begin{align}
	P(t'',x''|t',x')&=\frac{1}{\sqrt{(2\pi(t''-t))\det g(t')}}\\ \nonumber
	&\times\exp\left( -\frac{(t''-t')}{2}\sum_{i,j=1}^n\left[ \frac{x''-x'}{t''-t'}-f_i(x',t') \right]g_{ij}^{-1}(t')\left[ \frac{(x''-x')}{(t''-t')}-f_j(x',t') \right] \right).
\end{align}
The one-step $r=\frac{1}{2}$, or symmetric, approximate path integral formula for the transition probability amplitude is
\begin{align}
	&P(t'',x''|t',x')=\frac{1}{\sqrt{(2\pi(t''-t))\det g(\bar{t})}}\\ \nonumber
	&\times\exp\left( -\frac{(t''-t')}{2}\sum_{i,j=1}^n\left[ \frac{x''-x'}{t''-t'}-f_i(\bar{x},\bar{t}) \right]g_{ij}^{-1}(\bar{t})\left[ \frac{(x''-x')}{(t''-t')}-f_j(\bar{x},\bar{t}) \right] -\frac{(t''-t')}{2}\sum_{i=1}^n\frac{\partial f_i}{\partial x_i}(\bar{x},\bar{t})\right),
\end{align}
where $\bar{x}=\frac{1}{2}(x''+x')$ and $\bar{t}=\frac{1}{2}(t'+t'')$.

\begin{example}[Wiener Process]
	The simplest example is the following state model:
	\begin{align}
		d\rv{x}(t)=\sigma d\rv{v}(t).
	\end{align}
	The one-step formula for this case is 
	\begin{align}
		p(t'',x''|t',x')=\frac{1}{\sqrt{2\pi\sigma^2(t''-t')}}\exp\left( -\frac{(x''-x')^2}{2\sigma^2(t''-t')} \right).
	\end{align}
	This formula is well known to be exact for arbitrary size time step\cite{A.T.Bharucha-Reid1960}.  The extension to the $n-$dimensional case is straightforward.	
\end{example}

\begin{example}[Linear Model]
	Consider the following state model:
	\begin{align}
	d\rv{x}(t)=-0.2\rv{x}(t)dt+\sqrt{10}d\rv{v}(t).
	\end{align}
	The pre-point one-step formula is seen to be the lowest order approximation to the exact solution\cite{A.T.Bharucha-Reid1960}. Thus, this approximation is accurate for small time steps.
\end{example}

\begin{example}[A Polynomial Nonlinear Model]

Consider the model
\begin{align}\label{eq:ExPoly}
	d\rv{x}(t)=-4\rv{x}(t)(\rv{x}^2(t)-1)dt+0.1d\rv{v}(t).
\end{align}
The results for the transition probability density for $x'=-0.75,-0.35,-0.05,0.30,0.70$ are plotted in Figure \ref{fig:Ex3Poly} along with values obtained from $10^5$ simulations. The time step size is $0.1$. 

\FIGURE{
  \centering
  \scalebox{0.75}{ \includegraphics{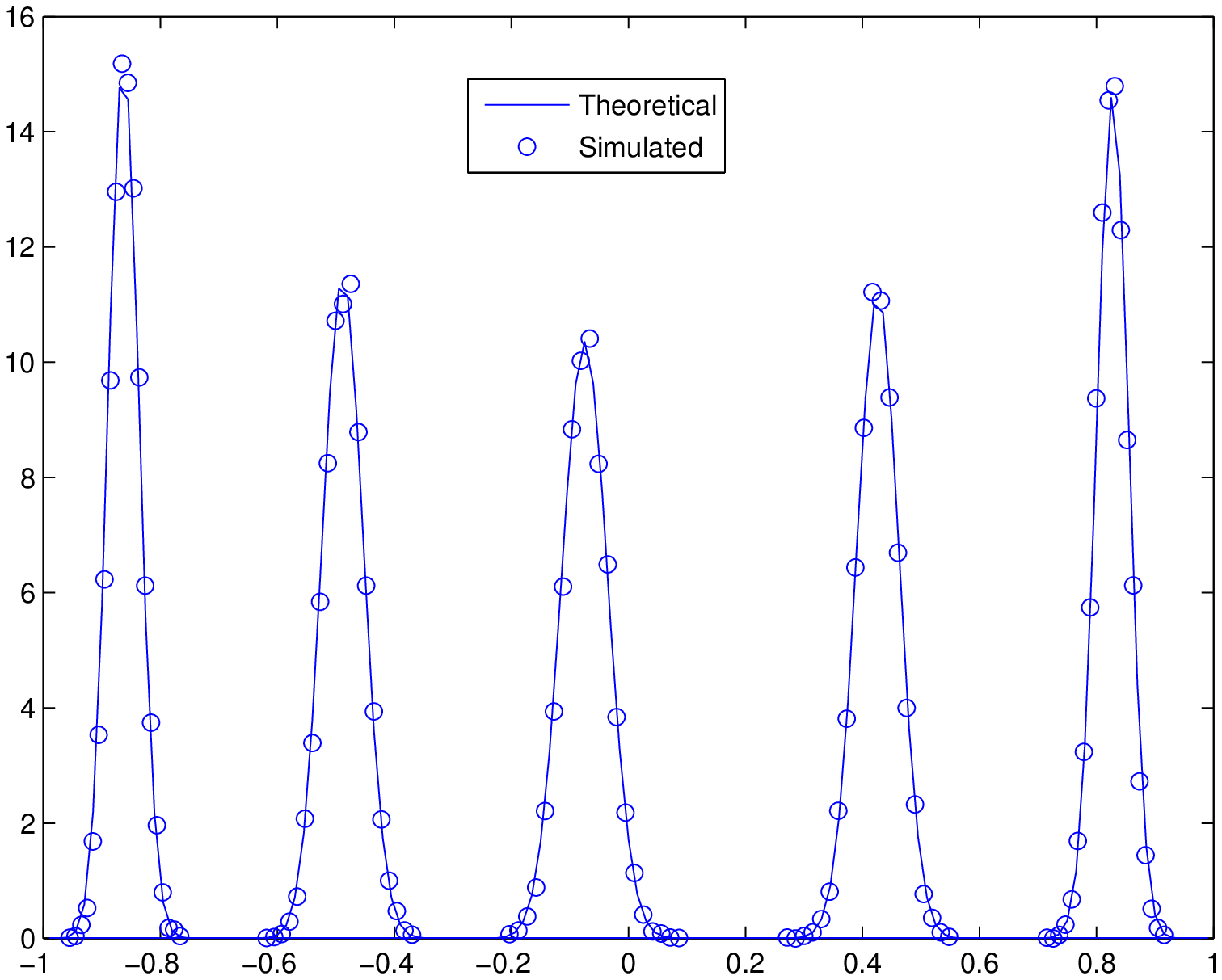}}
 	\caption{Transition  probability density for a nonlinear state model with polynomial drift (Equation \ref{eq:ExPoly}). Here $t''-t'=0.1$  and $x'=-0.75,-0.35,-0.05,0.30,0.70$.}
  \label{fig:Ex3Poly}
}

\end{example}

\begin{example}[A Transcendental Nonlinear Model]
	Consider the model
	\begin{align}\label{eq:Trans}
		d\rv{x}(t)=1.2\cos(3\rv{x}(t))dt+0.3d\rv{v}(t).
	\end{align}
	The symmetric one-step approximate path integral formula for the conditional probability density is plotted in Figure \ref{fig:Ex4Trans}. The time step size is 0.2. 

\FIGURE{
  \centering
  \scalebox{0.75}{ \includegraphics{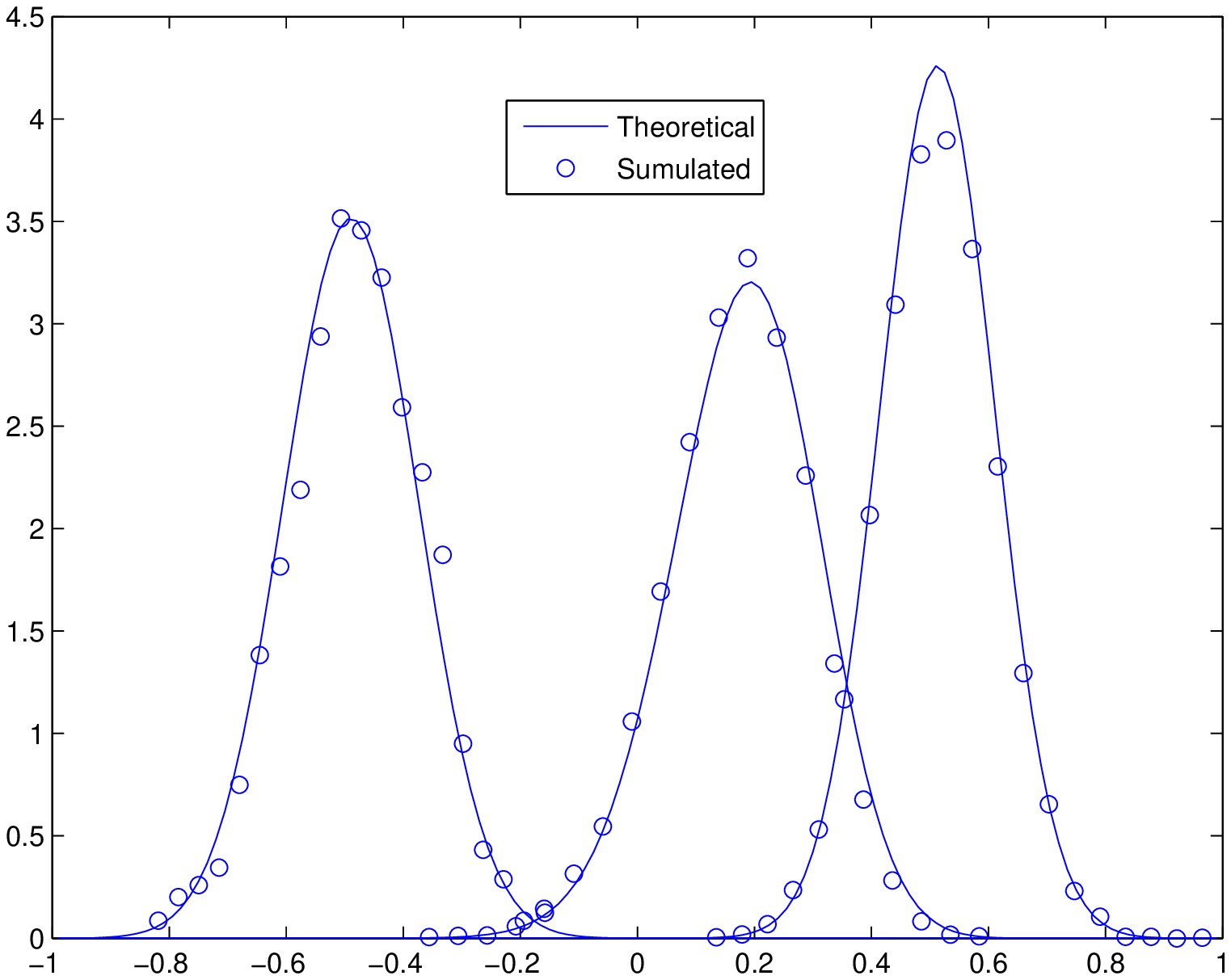}}
 \caption{Transition probability density for a nonlinear state model with transcendental drift (Equation {\ref{eq:Trans}}). Here $t''-t'=0.2$  and $x'=-0.5,-0.05,-0.05,0.5$.}
  \label{fig:Ex4Trans}
}

\end{example}
These examples illustrate that some of the simplest one-step path integral approximation formulas are often very accurate. Clearly, the smaller the time steps, the more accurate is the approximation. What is remarkable is  to note that \textit{the time step need not be infinitesimal for extremely accurate results}. In practice, the time steps are often small, and the one-step approximations discussed here might be adequate for many applications. For larger time steps, there are better one-step approximations that have been obtained and will yield better performance. 

Finally, observe that time dependence is not a problem provided the variation of the diffusion vielbein is small over the time interval $(t''-t')$. Of course, this means that the transition probability density depends on the actual times ($t''$ and $t'$),  rather than just their difference (as in the time-independent case).

\section{Additional Remarks}

\subsection{Physical Meaning of the Path Integral Formula}

A beautiful discussion of the path integral in the quantum mechanical context is given in \cite{R.P.FeynmanandA.R.Hibbs1965}.  Let us consider the model discussed in Section \ref{sec:pathinteg1}. Intuitively, the path integral is a sum of contributions over all paths satisfying the boundary conditions: $x(t_0)=x_0$ and $x(t)=x$. The action is computed for each path, and the exponential of the action is the contribution from that path. In other words, the contributions of the paths are exponentially weighted according to the action. It is clear that the dominant contribution comes from paths near the path of least action. 

This heuristic picture is seen to be consistent with intuition in the limit the noise is small. Let us consider the model discussed in Section \ref{sec:pathinteg1} and assume that the noise is absent. Then, the evolution of the state is governed by a deterministic  dynamical equation---the state evolves according to the zero noise dynamical law:
\begin{align}
	\label{eq:ODEZero}
	\dot{x}_i(t)=f_i(x(t)),\quad i=1,2,\dots,n.
\end{align}
According to the path integral formula, Equation \ref{eq:PIformula}, when the noise is small, action is small only if  the ordinary differential equation (ODE) Equation \ref{eq:ODEZero} is satisfied (or nearly so). This implies that, to a good approximation, the particle moves along the deterministic trajectory specified by Equation \ref{eq:ODEZero}, as one would intuitively expect. The path integral formula gives a precise sense in which the deviation from the deterministic path is probabilistically possible as the variance of the noise, $\hbar_{\nu}$, increases. This is analogous to the situation in quantum physics, where the role played by $\hbar_{\nu}$ is played by the Planck's constant $\hbar$.  

\subsection{Comments on the integral of S-T. Yau and S. S-T. Yau}

A solution in terms of ordinary integrals was presented in \cite{S.-T.YauS.S.-TYau1996}. The formal solution derived by S-T. Yau and S. S-T. Yau is of the form 
\begin{align}
	\label{eq:YauandYauKol}
	u(t,x)&=\int_{-\infty}^{\infty}\left\{ d^n\xi \right\}(2\pi t)^{n/2}\\ \nonumber
	&\times\exp\left( -\frac{1}{2t}\sum_{j=1}^n(x_j-y_j)^2+\int_0^1\sum_{i=1}^n(x_i-y_i)f_i(y+t(x,y))dt \right)\left[ 1+\sum_{i=1}^{\infty}\tilde{a_i}(x,y)t^i \right]\sigma_0(\xi),
\end{align}
for arbitrary initial condition $u(0,x)=\sigma_0(x)$ (for details see \cite{S.-T.YauS.S.-TYau1996}). The important thing to note is that this is an infinite series in $t$. In addition, they present an estimate of the time interval on which their solution converges. In contrast, the path integral solution is infinite-dimensional integral but with a much simpler structure\footnote{It is also important to note that they have represented an infinite dimensional path integral as a finite-dimensional ordinary integral. This may be useful in other areas, such as quantum mechanics and field theory.}.

Also, note that the result in \cite{S.-T.YauS.S.-TYau1996} differs from that obtained here in a few aspects. Firstly, they do not assume current conservation, and so solve a more general equation than the FPKfe. Secondly, the derivations here assume that the signal model noise  is additive, whereas in  \cite{S.-T.YauS.S.-TYau1996} a weaker condition is assumed, namely that the diffusion matrix is orthogonal.  However, unlike \cite{S.-T.YauS.S.-TYau1996} the FPKfe is solved for the explicit time-dependent case.

\section{Conclusion}

In this paper, the path integral formula has been derived for the Fokker-Planck-Kolmogorov forward equation that arises from the explicitly time-dependent Langevin dynamical equation with additive noise with a rectangular diffusion vielbein. This can be applied to solve the continuous-discrete filtering problems very accurately.

The path integral formulation has several advantages. It has been successfully used to solve challenging problems in quantum field theory not solvable using other approaches. Furthermore, from a conceptual point of view, the path integral formulation has proven to be extremely valuable. The insights from path integrals to problems in several other fields, especially quantum mechanics and field theory, has been tremendous.

Hence it is a very promising framework for studying the nonlinear filtering problem. Numerous numerical methods follow from the path integral formulas derived here. This is being utilized in subsequent work. 

\appendix

\renewcommand{\theequation}{\thesection.\arabic{equation}}
\section{ Gaussian Integrals}\label{sec:GaussianIntegrals}
It is remarkable that the foundation of the sophisticated methods of path integrals is based on Gaussian integrals. A Gaussian integral is defined as one where the integrand is an exponential function with an exponent quadratic  in the variables. The simplest Gaussian integral 
\begin{align}
	I=\int_{-\infty}^{\infty}e^{-x^2}dx,
\end{align}
can by evaluated by noting that 
\begin{align}
	I^2=\int_{-\infty}^{\infty}\int_{-\infty}^{\infty}e^{-(x^2+y^2)}dxdy.
\end{align}
Transforming to polar coordinates with the substitutions 
\begin{align}
	x&=r\cos\theta, \quad 0<r\le\infty,\\ \nonumber
	y&=r\sin\theta, \quad0<\theta\le2\pi,
\end{align}
so that the measure becomes $dxdy=rdrd\theta$, it follows that
\begin{align}
	I^2&=\int_0^{2\pi}d\theta\int_0^{\infty}e^{-r^2}rdr,\\ \nonumber
	&=2\pi\int_0^{\infty}e^{-z}\frac{dz}{2},\quad (z=r^2)\\ \nonumber
	&=\pi.
\end{align}
Hence,  
\begin{align}
	\int_{-\infty}^{\infty}e^{-x^2}dx=\sqrt{\pi},
\end{align}
or more generally (assuming $a>0$ so that the integral is convergent),
\begin{align}
	\int_{-\infty}^{\infty}e^{-ax^2}dx&=\int_{-\infty}^{\infty}e^{-z^2}\frac{dz}{\sqrt{a}},\quad z=\sqrt{a}x,\\ \nonumber
	&=\sqrt{\frac{\pi}{a}}. 
\end{align}
The general one-dimensional Gaussian integral can be evaluated by completing squares: 
\begin{align}
	\int_{-\infty}^{\infty}e^{-(ax^2+bx+c)}dx&=e^{-c}\int_{-\infty}^{\infty}e^{-a\left[ \left( x+\frac{b}{2a} \right)^2-\frac{b^2}{4a^2} \right]}dx,\\ \nonumber
	&=e^{\left( \frac{b^2}{4a}-c \right)}\int_{-\infty}^{\infty}e^{-z'^2}\frac{dz'}{\sqrt{a}},\quad z'=\sqrt{a}\left( x+\frac{b}{2a} \right),\\ \nonumber
	&=\sqrt{\frac{\pi}{a}}e^{\left(\frac{b^2}{4a} -c\right)}. 
\end{align}
Define the $n-$dimensional measure
\begin{align}
	\left\{ d^nx \right\}\equiv dx_1 dx_2\cdots dx_n.
\end{align}
Then, it follows that 
\begin{align}
	\int \left\{ d^nx \right\}e^{-\sum_{i=1}^na_ix_i^2}dx	
	&= \prod_{i=1}^n\int_{-\infty}^{\infty}dx_ie^{-a_ix_i^2},\\ \nonumber
	&=\sqrt{\frac{\pi^n}{\prod_{j=1}^na_j}},\quad a_i>0.
\end{align}
Next, consider the quadratic form
\begin{align}
	\alpha\sum_{i,j=1}^nx_iA_{ij}x_j,
\end{align}
where $\alpha$ is a positive real number and $A$ is a real $n\times n$ symmetric matrix\footnote{Note that any matrix can be written as a sum of a symmetric matrix($A^S$) and an antisymmetric matrix because of the identity 
\begin{align}
	A=\frac{A+A^T}{2}+\frac{A-A^T}{2}\equiv A^S+A^A,\quad A^S_{ij}=A^S_{ji}, \quad A^A_{ij}=-A^A_{ji}.
\end{align}
However, since $x_i$ are commuting variables, i.e., $x_ix_j=x_jx_i$, 
\begin{align}
	\sum_{i,j=1}^nx_iA^A_{ij}x_j&=\sum_{i,j=1}^nx_jA^A_{ji}x_i,\quad(i\leftrightarrow j)\\ \nonumber
	&=-\sum_{i,j=1}^nx_jA_{ij}^Ax_i =-\sum_{i,j=1}^nx_iA_{ij}^Ax_j=0.
\end{align}
Hence, without loss of generality, $A$ may be assumed to be a symmetric matrix. 
}. A real, symmetric matrix can always be diagonalized by an orthogonal transformation $O$:
\begin{align}
	A=O^T\Lambda(A) O,
\end{align}
Here, $\Lambda(A)$ is a diagonal matrix with its diagonal entries being the eigenvalues $\lambda_i(A)$ of $A$. Thus,
\begin{align}
	\alpha\sum_{i,j=1}^nx_iA_{ij}x_j&=\alpha\sum_{i,j=1}^ny_i\Lambda_{ij}(A)y_j,\\ \nonumber
	&=\alpha\sum_{i,j=1}^n\lambda_i(A)y_i^2,\quad y_i=\sum_{j=1}^nO_{ij}x_j.
\end{align}
Since the Jacobian is unity ($\det O=1$),  
\begin{align}
	\int \left\{ d^nx \right\}e^{-\alpha\sum_{i,j=1}^nx_iA_{ij}x_j}&=\int\left\{ d^ny \right\}e^{-\alpha \sum_{i=1}^n\lambda_i(A)y_i^2},\\ \nonumber
	&=\left( \frac{\pi}{\alpha} \right)^{n/2}\frac{1}{\sqrt{\prod_{i=1}^n\lambda_i(A)}},\\ \nonumber
	&=\left( \frac{\pi}{\alpha} \right)^{n/2}\frac{1}{\sqrt{\det A}},\quad\alpha>0.
\end{align}
The most general quadratic form can be written as follows:
\begin{align}
	\alpha\beta\sum_{i,j=1}^nx_iA_{ij}x_j+&\beta\sum_{i=1}^nb_ix_i+c=\\ \nonumber
	&\alpha\beta\sum_{i,j=1}^n\left( x_i+\frac{1}{2\alpha}\sum_{k=1}^n\left( A^{-1} \right)_{ik}b_k \right)A_{ij}\left(x_j+\frac{1}{2\alpha}\sum_{k=1}^n\left( A^{-1} \right)_{jk}b_k \right)\\ \nonumber
	&-\frac{\beta}{4\alpha}\sum_{i,j=1}^nb_i\left( A^{-1} \right)_{ij}b_j+c.
\end{align}
Under the substitution
\begin{align}
	y_i=x_i+\frac{1}{2\alpha}\sum_{k=1}^n\left( A^{-1} \right)_{ik}b_k,
\end{align}
$dy_i=dx_i$ and the limits of integration are unchanged. Therefore, the general $n-$dimensional Gaussian integral is 
\begin{align}\label{eq:Zdef}
	Z(A,b,c)&\equiv\int \left\{d^nx\right\}e^{-\beta\left[ \alpha\sum_{i,j=1}^nx_iA_{ij}x_j+\sum_{i=1}^nb_ix_i +c/\beta\right]},\\ \nonumber
	&=\int\left\{ d^ny \right\}e^{-\sum_{i,j=1}^ny_{i}(\alpha\beta A)_{ij}y_j+\frac{\beta}{4\alpha}\sum_{i,j=1}^nb_i\left( A^{-1} \right)_{ij}b_j-c},\\ \nonumber
	&=\left[\left( \frac{\pi}{\alpha\beta} \right)^{n/2}\frac{e^{-c}}{\sqrt{\det A}}\right]e^{\frac{\beta}{4\alpha}\sum_{i,j=1}^nb_iA_{ij}^{-1}b_j}.
\end{align}
The normalized measure
\begin{align}
	\label{eq:NormMeas}
	[d^nx]\equiv\left[ \sqrt{\det A}\left( \frac{\alpha\beta}{\pi} \right)^{n/2} \right],
\end{align}
is defined so that
\begin{align}
	\label{eq:NormMeasDef}
	\int[d^nx]e^{-\alpha\sum_{i,j=1}^nx_iA_{ij}x_j}&=\int\left\{ d^nx \left(\frac{\alpha\beta}{\pi} \right)^{n/2}\sqrt{\det A}\right\}e^{-\alpha\sum_{i,j=1}^nx_iA_{ij}x_j},\\ \nonumber
	&=1.
\end{align}

In the calculation of the expectation values of polynomials in random vectors with a Gaussian distribution, integrals of the form arise:
\begin{align}
	\int\left[ d^nx \right]x_{k_1}x_{k_2}\cdots x_{k_n}e^{-\alpha\sum_{i,j=1}^nx_iA_{ij}x_j}=\frac{\int \left\{ d^nx \right\} x_{k_1}x_{k_2}\cdots x_{k_l}e^{-\alpha\beta\sum_{i,j=1}^nx_iA_{ij}x_j}}{\int \left\{ d^nx \right\}e^{-\alpha\beta\sum_{i,j=1}^nx_iA_{ij}x_j}}.
\end{align}

From the Equation \ref{eq:Zdef}, it follows from repeated differentiation with respect to $\beta b$, that 
\begin{align}
	\vev{ \rv{x}_{k_1}\rv{x}_{k_2}\cdots\rv{x}_{k_l}}&=\left( \left( \frac{\alpha\beta}{\pi} \right)^{n/2}\sqrt{\det A} \right)\left[ \frac{(-1)^l}{\beta^l}\frac{\partial}{\partial b_{k_1}}\frac{\partial}{\partial b_{k_2}}\cdots\frac{\partial}{\partial b_{k_l}} Z(A,b,0) \right]\Big|_{b=0},\\ \nonumber
	&=\left( \frac{(-1)^l}{\beta^l}\frac{\partial}{\partial b_{k_1}}\frac{\partial}{\partial b_{k_2}}\cdots\frac{\partial}{\partial b_{k_l}}e^{\frac{\beta}{4\alpha}\sum_{i=1}^nb_i\left( A^{-1} \right)_{ij}b_j} \right)\Big|_{b=0}.
\end{align}
In fact, if $F(x_1,x_2,\ldots,x_n)$ is a power series in $x_i$:
\begin{align}
	\vev{F(\rv{x}_1,\rv{x}_2,\dots,\rv{x}_n)}=\left( F\left( -\frac{1}{\beta}\frac{\partial}{\partial b_1},-\frac{1}{\beta}\frac{\partial}{\partial b_2},\cdots,-\frac{1}{\beta}\frac{\partial}{\partial b_n} \right)e^{\frac{\beta}{4\alpha}\sum_{i,j=1}^nb_iA^{-1}_{ij}b_j} \right)\Big|_{b=0}.
\end{align}
An example is the two-point function:
\begin{align} 
	\label{eq:Two-PointFn}
	\vev{\rv{x}_i\rv{x}_j}&=\frac{(-1)^2}{\beta^2}\frac{\beta}{2\alpha}\left( A^{-1} \right)_{ij},\\ \nonumber
	&=\frac{1}{2\alpha\beta}\left( A^{-1} \right)_{ij}.
\end{align}

The expectation can be simplified further. Note that if $l$ is odd, the expectation vanishes. When $l$ is even, it is easy to show (using the method of induction) that:
\begin{align} 
	\vev{x_{k_1}x_{k_2}\cdots x_{k_l}}&=\sum_{\text{Permutations P of }\{k_1,k_2,\ldots,k_l\}}\left( \frac{\beta}{2\alpha} \right)^lA^{-1}_{k_{P_1}k_{P_2}}\cdots A^{-1}_{k_{P_l-1}k_{P_l}},\\ \nonumber
	&=\sum_{\text{Permutations of }\{k_1,k_2,\ldots,k_l\}}\left( \frac{\beta}{2\alpha} \right)^l\vev{x_{P_1}x_{k_{P_2}}}\cdots\vev{x_{k_{P_{l-1}}}x_{k_{P_l}}}.
\end{align}
This can be formally generalized to an infinite number of variables, i.e., fields, and is then referred to as the Wick's theorem for bosonic fields in quantum field theory.

\section{ Delta-Function and Constraints}\label{sec:DeltaFnConstraints}

The delta function\footnote{Strictly speaking, the delta function should be called the delta distribution defined rigorously as a limit of sequence of certain functions; it is not mathematically meaningful to view the Dirac delta function as a function. However, for most of our applications, the manipulations done here are valid if the delta function is always considered to be a part of the integrand.} has several representations. The most useful representation for our purposes is the following representation of the $n-$dimensional delta function
 \begin{align}\label{eq:deltadefn}
	\delta^n(x-x')
	&\equiv\frac{1}{(2\pi)^n}\int \left\{ d^nk \right\}e^{i\sum_{i=1}^nk_i(x_i-x'_i)},\\ \nonumber
	&\equiv\int\left[ d^nk \right]e^{i\sum_{i=1}^nk_i(x_i-x_i')}.
\end{align}
The Fourier transform and its inverse are defined as follows:
\begin{align}
	\tilde{f}(k)&=\int\left\{ d^nx \right\}e^{-i\sum_{i=1}^nk_ix_i}f(x), \\ \nonumber
	f(x)&=\frac{1}{(2\pi)^n}\int\left\{ d^nk \right\}e^{i\sum_{i=1}^nk_ix_i}\tilde{f}(k).
\end{align}

Consider the  problem of writing down the expression of a function $\sigma(x)$ at the point where $f(x)=0$ without actually solving the constraint equation. Assume that the solution to the equation $f(x)=0$ is unique. Then,
\begin{align}
	\sigma(x)|_{f(x)=0}&=\sigma(x_c),\quad f(x_c)=0,\\ \nonumber
	&=\int \left\{ d^nx \right\}\delta(x-x_c)\sigma(x).
\end{align}
From the sequence of  identities
\begin{align}\label{eq:UnityFn}
	1&=\int \left\{ d^ny \right\}\delta^n(y),\\ \nonumber
	&=\int \left\{ d^nf(x) \right\}\delta^n(f(x)),\\ \nonumber
	&=\int\left\{  d^nx \right\} J(x)\delta^n(f(x)),\quad J=\det\left( \frac{\partial f_i}{\partial x_j}(x) \right),\\ \nonumber
	&=\int \left\{  d^nx \right\}\delta^n(x-x_c),
\end{align}
it follows that
\begin{align}
	\delta^n(f(x))&=\frac{\delta^n(x-x_c)}{J(x)},\\ \nonumber
	\delta^n(x-x_c)&=\delta^n(f(x))J(x).
\end{align}
Hence, 
\begin{align}
	\sigma(x)&=\int \left\{  d^nx \right\}\delta^n(f(x))J(x)\sigma(x).
\end{align}
Using the Fourier integral representation of the delta function in Equation \ref{eq:deltadefn}, it is clear that 
\begin{align}
	\sigma(x)|_{f(x)=0}&=\frac{1}{(2\pi)^n}\int \left\{  d^nx \right\} \left\{  d^n\lambda \right\} e^{i\lambda f(x)}J(x)\sigma(x). 
\end{align}

\section{Functional Calculus: A Brief Note}\label{sec:FuncCalc}

Recall that a function $f(x)$ gives a number for each point $x$. The infinite dimensional generalization of a function is called a functional, $f[x(t)]$, which gives a number for each function $x(t)$. 
Loosely speaking, the functional derivative may be defined analogous to the ordinary derivative as follows:
\begin{align}
	\frac{\delta f(x(t))}{\delta x(t')}=\lim_{\epsilon\ra0}\frac{f(x(t)+\epsilon\delta(t-t'))-f(x(t))}{\epsilon}.
\end{align}
Alternatively, the functional derivative, $\frac{\delta f}{\delta x(t')}(x(t))$, of the functional $f(x(t))$ with respect to variation of the function $x(t)$ at $s$ is defined  by the following equation
\begin{align} 
	f[x(t)+\eta(t))=f[x(t))+\int\frac{\delta f}{\delta x(s')}(x(s))\eta(s')ds'+\cdots. 
\end{align}
Note that the functional derivative of a functional is also a functional. Functional differentiation obeys the standard algebraic rules (linearity and Leibnitz's rule):
\begin{align}
	\frac{\delta}{\delta x(t')}\left[ \sum_{i=1}^nf_{i}(x(t)) \right]&=\sum_{i=1}^n\frac{\delta f_i}{\delta x(t')}(x(t)),\\
	\frac{\delta}{\delta x(t')}\left[f_1(x(t))f_2(x(t))\right]&=f_1(x(t))\frac{\delta}{\delta x(t')}f_2(x(t))+f_2(x(t))\frac{\delta}{\delta x(t')}f_1(x(t)).
\end{align}
Just as the derivative of $x$ with respect to $x$ is 1, the functional derivative of $x(t)$ with respect to $x(t')$ is the unit matrix in infinite dimensions, namely the delta function:
\begin{align}
	\frac{\delta x(t)}{\delta x(t')}=\delta(t-t').
\end{align}
The functional derivatives of a formal power series in function $x(t)$ can be seen to be analogous to the ordinary derivative of a power series in the variable $t$. Specifically, it is straightforward to verify that if 
\begin{align}
	f\left( x(t) \right)= \sum_{n=0}^{\infty}\frac{1}{n!}\int dt_1\cdots dt_n f^{(n)}(t_1,t_2,\dots,t_n)x(t_1)\cdots x(t_n),
\end{align}
then 
\begin{align}
	f^{(r)}(t_1,t_2,\dots,t_r)=\left\{ \left( \prod_{i=1}^r\frac{\delta}{\delta x(t_i)} \right)f \right\}\Bigg|_{x=0}.
\end{align}
In particular, a functional representable as a power series in functions, like the exponential functional, can be functionally differentiated using this result. Thus
\begin{align}
	\frac{\delta}{\delta J(x)}\exp\left( \int dx'J(x')A(x') \right)=A(x)\exp\left( \int dx'J(x')A(x') \right).
\end{align}
The functional delta function may be written as
\begin{align}
	\label{eq:FuncDeltaFn}
	\delta(f(x(t))=\int\left[ d\lambda(t) \right]\exp\left(i\int dt\lambda(t)f(x(t))  \right).
\end{align}
Note that Equation \ref{eq:UnityFn} becomes
\begin{align}
\label{eq:UnityFunctional}
	1=\int[\mc{D}f(x(t')))\det\left[ \frac{\delta f}{\delta x(t')}(x(t))\right] \delta(f(x(t))).
\end{align}

In the application at hand, determinants of operators (matrices) arise in Gaussian integration in the infinite (finite) dimesional case. Of particular interest is the operator $\det(\delta(x-y)+K(x,y))$, for some operator $K(x,y)$, where it is assumed that  traces of all powers of $K$ exist. From the identity
\begin{align}
	\text{ln}\det M=\text{tr}\text{ln}M,
\end{align}
it follows that
\begin{align}\label{eq:DetIdentity}
	\ln\det[1+K]=&\int dxK(x,x)-\frac{1}{2}\int dx_1dx_2K(x_1,x_2)K(x_2,x_3)+\cdots+\\ \nonumber
	&\frac{(-1)^{n+1}}{n}\int dx_1\cdots dx_n K(x_1,x_2)K(x_2,x_3)\cdots K(x_n,x_1)+\cdots.
\end{align}
Note that if $K(x,y)=\theta(x-y)\tilde{K}(x,y)$, only the first term is nonzero.

\bibliographystyle{IEEEtran.bst}
\bibliography{onfbib}

\end{document}